\documentclass[fleqn,usenatbib]{mnras}

\usepackage{newtxtext,newtxmath}
\usepackage{CJKutf8}
\usepackage[T1]{fontenc}

\DeclareRobustCommand{\VAN}[3]{#2}
\let\VANthebibliography\thebibliography
\def\thebibliography{\DeclareRobustCommand{\VAN}[3]{##3}\VANthebibliography}

\usepackage{xcolor}

\usepackage{graphicx}	
\usepackage{amsmath}	

\title[Transport-induced Chemistry on K2-18b, Part I]{Three-dimensional Transport-induced Chemistry on Temperate sub-Neptune K2-18b, Part I: the Effects of Atmospheric Dynamics}

\author[J. Liu et al.]{
\begin{CJK*}{UTF8}{gbsn}Jiachen Liu (刘伽晨),$^{1,2,3}$\thanks{E-mail: jiachenliu@mpia.de}
Duncan Christie$^{1,3}$ and
Jun Yang (杨军)$^{2}$\end{CJK*} 
\\
$^{1}$Max Planck Institute for Astronomy, Heidelberg, 69117, Germany\\
$^{2}$Department of Atmospheric and Oceanic Sciences, School of Physics, Peking University, Beijing 100871, People's Republic of China\\
$^{3}$Department of Physics and Astronomy, Faculty of Environment, Science and Economy, University of Exeter, Exeter EX4 4QL, UK
}

\date{Accepted XXX. Received YYY; in original form ZZZ}

\pubyear{\the\year{}}

\begin{document}

\label{firstpage}
\pagerange{\pageref{firstpage}--\pageref{lastpage}}
\maketitle

\begin{abstract}
The low equilibrium temperatures of temperate sub-Neptunes lead to extremely long chemical timescales in their upper atmospheres, causing the abundances of chemical species to be strongly shaped by atmospheric transport. Here, we used a three-dimensional (3D) general circulation model involving a passive tracer to investigate the atmospheric circulation and 3D transport of temperate gas-rich sub-Neptunes, using K2-18b as an example. We model K2-18b as a synchronous or asynchronous rotator, exploring spin-orbit resonances (SOR) of 2:1, 6:1, and 10:1. We find that the strong absorption of CO$_2$ and CH$_4$ induces a detached convective zone between 1 and 5~bar, resulting in strong vertical mixing at these levels. The upper atmosphere is dominated by eastward winds (an equatorial superrotating jet present in all simulations), leading to warmer evening terminators and approximately 20$\%$ higher passive tracer mass mixing ratios compared to the morning terminators. Rotation rates have minimal impact on the strength of global mean vertical mixing, but significantly influence the latitudinal distribution of passive tracers. For synchronous, 2:1 SOR, and 6:1 SOR simulations, passive tracers are more abundant in the upwelling branches at latitudes within 60$^\circ$, while for the 10:1 SOR simulation, strong transient eddies at high latitudes ($>$70$^\circ$) between 0.1 to 1~bar can transport passive tracers upward from the deep atmosphere, making them more abundant there, despite their alignment with the downwelling branch of the large-scale circulation. This study focuses on the atmospheric dynamics and its influence on passive tracer transport, while a follow-up paper will incorporate active chemical species.

\end{abstract}




\section{Introduction}\label{sec:intro}

Detection of exoplanets has revealed that sub-Neptunes, with radii between 1.7 to 3.5 times Earth's radius, are abundant in the Galaxy \citep{batalha2014exploring,fulton2018california}. Among them, temperate sub-Neptunes (with equilibrium temperatures between 250 and 500 K) have gained significant interest due to their possible clear-sky conditions facilitating the detection of molecules and their potentially habitable environments. Previous studies suggest that temperate sub-Neptunes with equilibrium temperatures below 500~K should be less cloudy \citep{yu2021haze,dymont2022cleaning,brande_clouds_2024}, making them high-priority targets for observation. 

Sub-Neptunes are generally thought to have an accreted primordial H$_2$/He atmosphere with solid mantles and cores, akin to a smaller version of the gas dwarf planets \citep[e.g.,][]{misener2021cool}. However, their bulk density might also be explained by `water world' scenario with substantial H$_2$O in the condensed or steam form \citep[e.g.,][]{Zeng_planet_distribution_2019,luque_density_2022}. Given a sufficiently low surface temperature, recent studies propose that these H$_2$O-rich temperate sub-Neptunes could have a liquid ocean surface, categorized as `Hycean worlds' \citep[e.g.,][]{madhusudhan_interior_2020,Piette_Madhu_2020,Nixon_Madhu_ocean_2021}, providing a suitable environment for harboring life.

Up to now, JWST has detected molecules on two temperate sub-Neptunes: K2-18b \citep{madhusudhan_carbon-bearing_2023,madhusudhan2025new} and TOI-270d \citep{benneke2024jwst,holmberg2024possible}. Of them, K2-18b has gained significant interest as it receives a similar stellar flux as Earth ($\sim$1367 W\,m$^{-2}$), located in the canonical habitable zone \citep{kasting_habitable_1993,kopparapu_habitable_2013}. K2-18b has a mass of about 8.6 M$_\oplus$, a radius of about 2.6 R$_\oplus$, and an orbital period of 32.94 Earth days\footnote{Unless otherwise specified, `day' hereafter refers to an Earth day.} \citep{Montet_K2_2015,benneke_water_2019}. It has an equilibrium temperature of $\sim$280~K with zero planetary albedo. \citet{madhusudhan_carbon-bearing_2023} reported a detection of $\sim$1$\%$ volume mixing ratio of CH$_4$ and $\sim$1$\%$ volume mixing ratio of CO$_2$. The featured JWST transmission spectrum of K2-18b provides critical insights into its atmospheric composition and interior structure, while also serving as a key constraint for forward atmospheric modeling.

\citet{madhusudhan_carbon-bearing_2023} argued that the detection of CH$_4$ and CO$_2$, along with the non-detection of NH$_3$, provides evidence that K2-18b could be a Hycean world, as a water ocean can lead to the depletion of NH$_3$ \citep{Hu_solubility_nh3_2021}. However, this conclusion is later disputed by \citet{shorttle_distinguishing_2024}, showing that the high solubility of NH$_3$ in magma ocean can explain its depletion. Moreover, subsequent photochemical modeling by \citet{wogan2024jwst} suggested that the lifeless Hycean world scenario does not agree with the data. Instead, either an `inhabited' Hycean world with a biological CH$_4$ flux or a gas-rich mini-Neptune\footnote{`Sub-Neptune' is used to classify planets based on size (analogous to super-Earths) while `mini-Neptune' refers to planets categorized by their interior structure (analogous to Hycean world). A thorough discussion of sub-Neptunes' interior and envelope structure can be found in \citet{benneke2024jwst}.} with 100 times solar metallicity can provide a comparable agreement to the observations. \citet{wogan2024jwst} argued that the mini-Neptune scenario is more favorable, as it is simpler and does not require the presence of a biosphere.  Later, \citet{cooke2024considerations} reexamined the three scenarios considered by \citet{wogan2024jwst} and rebutted that the inhabited Hycean world scenario should be preferred over the mini-Neptune scenario, as its chemical abundances are consistent with all retrieved values in \citet{madhusudhan_carbon-bearing_2023}. A recent study reported hints of possible signatures on K2-18b using JWST MIRI observation, but has evoked debates \citep{seager_prospects_nodate,taylor_are_2025,welbanks_challenges_2025,luque_insufficient_2025}.

Due to the limited transmission spectra and relatively low signal-to-noise ratio of the JWST data, the debate on K2-18b's interior is not resolved yet. In this study, we assume that K2-18b is a mini-Neptune. While we do not explore the Hycean world or magma ocean scenario, a better understanding of K2-18b as a gas-rich mini-Neptune may help resolve the degeneracy in its structural interpretation. We note that the current study was initiated when the observational data from \citet{madhusudhan_carbon-bearing_2023} were the primary reference available, so the simulation setup (e.g., metallicity, cloud-free and haze-free scenario) was made according to this paper. We acknowledge that a recent reanalysis by \citet{schmidt2025comprehensive} casted doubt on the robustness of the CO$_2$ detection reported in \citet{madhusudhan_carbon-bearing_2023}, and suggests that an oxygen-poor mini-Neptune model may also be consistent with the observations. Additional JWST transmission spectra for K2-18b are anticipated in the near future \citep{Hu_JWST_2021}, which will provide improved constraints and further refine our understanding of this intriguing planet.

One common way to interpret the atmospheric composition from observational data is to do forward chemical modelling. The low equilibrium temperature of mini-Neptune K2-18b leads to exceptionally long chemical timescales ($\tau_\mathrm{chem}\gg 10^{10}$ s) compared to the vertical mixing timescale ($\tau_\mathrm{mix}$) in the cold upper atmosphere \citep{Visscher_2011,Moses_2011}. As a result, the chemical abundances are strongly influenced by atmospheric transport and deviate from chemical equilibrium (i.e., $\tau_\mathrm{chem} > \tau_\mathrm{mix}$). The abundance of a molecular constituent may become `quenched' at a value representative of the quench level, which is defined by $\tau_\mathrm{chem} = \tau_\mathrm{mix}$.  
This can result in a significant change in molecular abundances in the photosphere, a region between $\sim$10$^{-2}$ and $\sim$10$^{-4}$~bar, which impacts the observed transmission spectrum. Furthermore, horizontal mixing, particularly driven by the strong jets in the upper atmosphere, can influence the horizontal distribution of chemical species, potentially affecting the observed features at the morning and evening terminators \citep[e.g.,][]{agundez2012impact,agundez2014pseudo,Baeyens_2021}. In light of these, understanding the effects of atmospheric dynamics on the three-dimensional transport of chemical species is important to correctly interpret the observation data.

Previous studies on the disequilibrium chemistry of K2-18b as a mini-Neptune have predominantly used one-dimensional (1D) models \citep[e.g.,][]{blain_1d_2021,tsai_inferring_2021,yu__how_2021,bezard_methane_2022,wogan2024jwst,cooke2024considerations}. These models parameterize vertical mixing using the eddy diffusion coefficient ($K_{zz}$). However, this approach oversimplifies the inherently complex nature of vertical mixing, and the values of $K_{zz}$ are often poorly constrained, reducing the reliability of the results. Furthermore, exoplanetary atmospheres are intrinsically three-dimensional (3D), with atmospheric dynamics influencing both vertical and horizontal transport of chemical species. Since 1D models do not account for horizontal mixing, they struggle to fully capture the complexity of transport-induced chemistry on K2-18b.

So far, three studies have conducted 3D simulations of K2-18b \citep{charnay2021formation,innes2022atmospheric,Barrier_convection_2025}. \citet{charnay2021formation} investigated cloud formation on K2-18b using a model with a non-gray radiative scheme and cloud scheme. Their results suggest that, assuming K2-18b is a synchronous rotator, its upper atmosphere is primarily shaped by a symmetric day-night overturning circulation, with clouds preferentially forming at the substellar point or along the terminators. \citet{innes2022atmospheric} employed a double-gray radiative scheme to study the dry atmospheric circulation, finding that K2-18b's upper atmosphere is dominated by eastward winds and an equatorial superrotating jet. \citet{Barrier_convection_2025} adopted a new convection scheme to model the convection on K2-18b. However, none of these studies focused on atmospheric transport or transport-induced chemistry, leaving these aspects largely unexplored.

In this study, we present the first three-dimensional investigation of transport-induced disequilibrium chemistry in the atmosphere of the temperate mini-Neptune K2-18b. Since atmospheric dynamics play a crucial role in shaping the distribution of chemical species, understanding these processes is essential for accurately interpreting the planet’s atmospheric composition. To address this, we structure our work into two consecutive papers.

Part I (this paper) focuses on characterizing the atmospheric circulation and its influence on three-dimensional transport in simulations that include a fully coupled chemical kinetics scheme. To isolate the role of atmospheric dynamics, we use a passive tracer that is advected by atmospheric motions but does not interact with radiation or chemical reactions. Although we do not analyse the chemical composition in this paper, we emphasize that the simulations do incorporate full chemical kinetics.

In Part II, we analyse the resulting chemical structure, derive an equivalent $K_{zz}$ profile for further use in 1D models, and compare the synthetic transmission spectra with JWST observations reported by \citet{madhusudhan_carbon-bearing_2023}.

The outline of this paper is as follows. In section~\ref{sec:methods}, we introduce the GCM used in this study and describe the experimental design. In section~\ref{subsec:thermal} and \ref{subsec:circulation}, we discuss the thermal structure and the atmospheric circulation on K2-18b, while in section~\ref{subsec:tracer}, we describe and explain the passive tracer distribution. Finally, conclusions and discussion are given in section~\ref{conclusions}.

\section{Model Description and Experimental Design} \label{sec:methods}

\subsection{Unified Model}\label{subsec:UM}
The 3D GCM we employed in this study is the Met Office Unified Model (UM). UM was originally developed to model Earth's atmosphere, then modified to simulate hydrogen-dominated hot Jupiters \citep[e.g.,][]{mayne2014unified,amundsen2016uk,drummond_effects_2016,mayne2017results,drummond_observable_2018,drummond2020implications,zamyatina_observability_2022,christie_impact_2021} and sub-Neptunes \citep[e.g.,][]{drummond_effect_2018,mayne2019limits,christie_impact_2022}. The model setup for a hydrogen-dominated atmosphere is comprehensively described in \citet{mayne2014unified, mayne2017results}.

The dynamic core of UM solves the non-hydrostatic equations of motion using a semi-implicit semi-Lagrangian scheme using finite difference on an Arakawa C grid \citep{wood2014inherently}. The model employs a geometric height-based grid for the vertical domain. Gravity varies with height, following $g(r) = g_\mathrm{p}(R_\mathrm{p}/r)^2$, where $g_\mathrm{p}$ and $R_\mathrm{p}$ are the surface gravity and planetary radius, and $r$ is the radial distance.
The horizontal resolution used in this study is 2.5$^{\circ}$$\times$2$^{\circ}$ in longitude versus latitude. The pressure level ranges from 200~bar to $\sim$1 Pa, with 66 vertical levels equally spaced in height. The radiation is calculated by a two-stream correlated-k radiative transfer model SOCRATES \citep{edwards1996studies,edwards1996efficient}, with 32 spectral bands covering 0.2-322 $\mu$m. We include the opacity of H$_2$O, CH$_4$, NH$_3$, CO$_2$, and CO, the H$_2$-H$_2$ collision-induced absorption, H$_2$-He collision-induced absorption, and the Rayleigh scattering due to H$_2$ and He. 

The model employs three methods to obtain stability: horizontal diffusion on the zonal wind, vertical damping at the boundary levels, and a bottom drag (See Appendix \ref{stability} for more information). These methods can ensure the model runs stably for at least ten thousand days, with the angular momentum loss within 3 percent and the total mass loss within 1.5 percent (Fig.~\ref{fig:conservation}).

\subsection{K2-18b set up}\label{subsec:K2-18b set up}

\begin{table}
	\centering
	\caption{Parameters used in this study.}
	\label{k2-18b}
	\begin{tabular}{cc} 
		\hline
		\textbf{Parameter} & \textbf{K2-18b}\\
		\hline
		Stellar flux & 1367~W\,m$^{-2}$  \\
            Inner radius ($R_\mathrm{p}$)& 16,430~km \\
            Orbital period & 32.94~days \\
            Planetary mass & 5.15$\times$10$^{25}$~kg \\
            Semi-major axis & 0.1591~au \\
            Surface gravity ($g_\mathrm{p}$) & 12.44~m\,s$^{-2}$ \\  
            Atmospheric composition & 180 $\times$ solar metallicity\\
            Specific gas constant ($R$)& 1023~J\,kg$^{-1}$\,K$^{-1}$ \\
            Specific heat capacity ($c_p$) & 3733~J\,kg$^{-1}$\,K$^{-1}$ \\
            Mean molecular weight & 8.46~g\,mol$^{-1}$ \\
            Internal temperature & 60~K (3.7~W\,m$^{-2}$) \\
            Domain height & 800~km \\
            Bottom and upper pressure & 200 and $\sim$10$^{-5}$~bar \\
            Horizontal resolution & 2.5$^{\circ}$$\times$2$^{\circ}$\\
		\hline
	\end{tabular}
\end{table}

K2-18b is a sub-Neptune with a mass of about 8.6 M$_\oplus$, a radius of about 2.6 R$_\oplus$, and an orbital period of 32.94 days \citep{benneke_water_2019}. It receives a similar stellar flux as Earth ($\sim$1367 W\,m$^{-2}$) and has an equilibrium temperature of $\sim$280~K assuming zero planetary albedo. The parameters of this planet are listed in Table \ref{k2-18b}. 

Considering the long orbital period of K2-18b, we compare its tidal spin-down timescales with the age of the K2-18 system to assess whether the planet might be an asynchronous rotator. The tidal spin-down timescale of the planet is given by \citet{guillot1996giant}:
\begin{equation}
    \tau_\mathrm{t} = Q \left( \frac{R^3_\mathrm{p}}{GM_\mathrm{p}}\right)\omega_\mathrm{p}\left(\frac{M_\mathrm{p}}{M_*}\right)^2\left(\frac{D}{R_\mathrm{p}}\right)^6,
\end{equation}
in which $Q$ is the planet's tidal dissipation rate, $G$ is the gravitational constant, $R_\mathrm{p}$ and $M_\mathrm{p}$ are the planet's radius and mass, $M_*$ is the mass of the star, $\omega_\mathrm{p}$ is the planet's primordial rotation rate, and $D$ is the semi-major axis. Assuming a tidal dissipation factor for sub-Neptunes, $Q=10^4$ \citep{louden2023tidal} and a primordial rotation rate similar to Jupiter's, $\omega_\mathrm{p}=1.7\times10^{-4}$~s$^{-1}$ \citep{guillot1996giant}, $\tau_\mathrm{t}$ is estimated to be approximately 1 Gyr. The age of K2-18 system ($\tau_\mathrm{a}$) estimated using the relation in \citet{engle2018rotation} is approximately 2.37 Gyr, and is estimated to be 2.9-3.1 Gyr based on gyrochronology in \citet{Sairam2025arXiv250319908S}.

Given the comparable magnitude of $\tau_t$ and $\tau_a$, K2-18b likely has not had sufficient time to become fully 1:1 tidally locked. In light of this, we investigate both synchronous and asynchronous configurations. To explore the latter, we model spin-orbit resonances (SOR) at 2:1, 6:1, and 10:1, corresponding to rotation periods of 16.47, 5.49, and 3.29 days, respectively. The asynchronous scenario of K2-18b is also considered and modeled in \citet{charnay2021formation}. In this study, both obliquity and eccentricity are assumed to be zero for simplicity. Nevertheless, \citet{cloutier2017characterization} estimated the eccentricity of K2-18b to be approximately 0.2, while \citet{Sarkis_2018} reported it to be less than 0.4.

Because the spectrum of K2-18 (effective temperature of 3,457~K) has not been measured, we instead use the spectrum of GJ 176, an M2.5 star with an effective temperature of 3,670~K, measured by MUSCLES survey \citep{france2016muscles}. For all the simulations, we set the internal temperature to 60~K, corresponding to an internal heat source of 3.7~W\,m$^{-2}$, following \citet{charnay2021formation}. 

Clouds and hazes are not included in this study. However, as noted by \citet{madhusudhan_carbon-bearing_2023} and \citet{wogan2024jwst}, a cloud-free and haze-free scenario provides a good agreement to the JWST observations of K2-18b reported in \citet{madhusudhan_carbon-bearing_2023}. This supports the assumption that neglecting clouds and hazes in this study should be reasonable. 

Considering the long convergence times of sub-Neptunes' 3D simulations noted by \citet{wang2020extremely} and the substantial computational cost of the chemical kinetics scheme, we do the simulations in two steps. First, we spin up the model by prescribing molecular abundances for the chemical species that affect opacity (referred to as the `fixed abundance run') until the system reaches a dynamical equilibrium state, discussed below. After the fixed abundance runs reach dynamical equilibrium, the model is restarted with chemical kinetics enabled (`kinetics run'), which involves chemical reactions, advection of chemical species by atmospheric circulation, and the interaction between chemical species and radiation. 

The chemical kinetics scheme employed in this study solves a set of ordinary differential equations (ODEs) to describe the production and loss of chemical species, based on a chemical network. These ODEs are integrated using the DLSODES solver \citep{hindmarsh1983scientific}. The chemical network is adapted from \citet{venot2019reduced}, consisting of 30 chemical species and 181 reversible reactions, excluding photodissociation processes. This scheme has been previously applied in studies of hot gas dwarfs \citep{drummond2020implications, zamyatina_observability_2022, zamyatina_quenching-driven_2024}.
A detailed discussion of the chemical kinetics scheme and its implications can be found in \citet{drummond2020implications} and \citet{zamyatina_observability_2022, zamyatina_quenching-driven_2024}, and will be presented in Part II, where we analyse the chemical structure. In addition to chemical species, a passive tracer that is advected by the flow but not actively affected by chemical reactions or interacts with radiation is incorporated into the kinetics runs to diagnose the effect of atmospheric transport. The setup of the passive tracer will be discussed in section~\ref{subsec:tracers set up}.

The fixed abundance runs are initialized from rest, using a temperature-pressure profile from the 1D radiative-convective equilibrium model \texttt{ATMO} assuming chemical equilibrium \citep{tremblin_fingering_2015,tremblin2016cloudless,drummond_effects_2016}. This \texttt{ATMO} simulation assumes 180 times solar metallicity, a solar C/O ratio (0.55), and effective heat redistribution, corresponding to a geometry factor of 0.25. Molecular abundances used in the fixed abundance runs are derived from \texttt{ATMO} under chemical kinetics, with the same metallicity and C/O ratio, assuming a $K_{zz}$ of 10$^6$ cm$^2$\,s$^{-1}$ and excluding photochemistry. The above-chosen metallicity and $K_{zz}$ values are based on the best agreement between the synthetic transmission spectrum from this simulation and JWST observations reported in \citet{madhusudhan_carbon-bearing_2023} among the reasonably selected parameters (Fig.~\ref{STS_1D}). For a more detailed discussion of these parameter choices, please refer to Appendix \ref{1D}, 
and the results from the chemical kinetics \texttt{ATMO} simulations can be seen in Fig.~\ref{ATMO_mole}.
The mean molecular weight, specific heat capacity, and specific gas constant used in the 3D simulations are derived from the vertical mean value from this 1D chemical kinetics simulation (shown in Table \ref{k2-18b} and Figs \ref{Cp_R}(a) and (b)). 
Equilibrium of the fixed abundance run is identified when the kinetic energy stabilizes, or the top-of-atmosphere energy imbalance is within 3~W\,m$^{-2}$.  The fixed abundance runs are run for 5,500 days for the synchronous and 2:1 SOR simulations, 8,000 days for the 6:1 SOR simulation, and 10,000 days for the 10:1 SOR simulation (Fig.~\ref{fig:convergence}). 

The kinetics runs are initialized with the 3D temperature and wind speed fields obtained from the preceding fixed abundance runs. The chemical abundances are initialized with chemical equilibrium abundances calculated by the initial 3D temperature fields. In the kinetics runs, the metallicity is set to 180 times solar, assuming the solar C/O ratio, while the other parameters remain the same as the fixed abundance runs. The kinetics runs are then run for an additional 5,100 days (approximately 155 K2-18b years), which is sufficient for molecular abundances in the photosphere to stabilize to a quasi-steady state (Fig.~\ref{fig:evolution}). 
The final 330 days of the kinetics runs (corresponding to 10 K2-18b years) are averaged at an output frequency of 10 days for subsequent analysis. 

The fixed abundance run requires approximately 1 day for 900 days of simulation with 864 CPUs on the Max Planck Society's Viper High-Performance Computing system. The kinetics run requires about two to three times the wall time of the fixed abundance run.

\subsection{Passive tracer parameterization}\label{subsec:tracers set up}

As discussed in section~\ref{sec:intro}, the cool temperature in the upper atmosphere leads to extremely long chemical timescales in the upper atmosphere, making the abundance and distribution of chemical species significantly affected by atmospheric transport. We begin by comparing the chemical timescale ($\tau_\mathrm{chem}$), the vertical mixing timescale ($\tau_\mathrm{mix}$), and the horizontal advection timescale ($\tau_\mathrm{adv}$) to gain a first-order understanding of which atmospheric layers are dominated by atmospheric transport. This comparison also informs the setup of the passive tracer.

\begin{figure}
\centering
\includegraphics[width=\columnwidth]{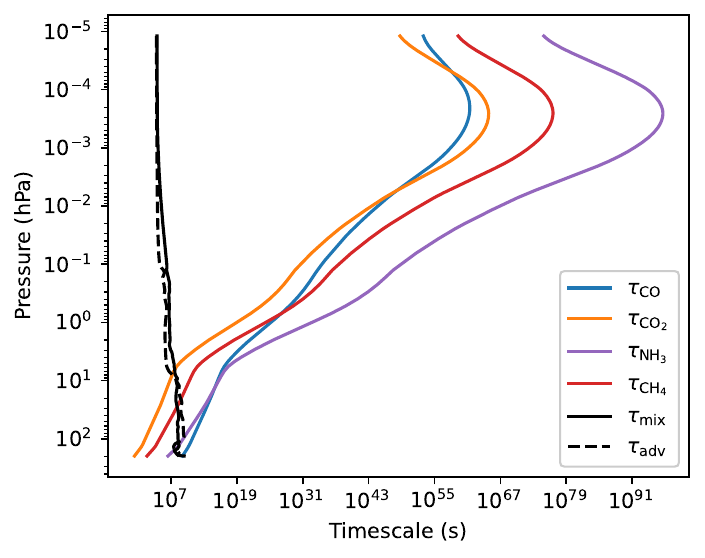}
\caption{The comparison between $\tau_\mathrm{chem}$ of CO, CO$_2$, NH$_3$, and CH$_4$ with $\tau_\mathrm{mix}$ and $\tau_\mathrm{adv}$. Chemical timescales are calculated using the Arrhenius-like fits provided in \citet{zahnle_methane_2014}, where they are expressed as: $\tau_{\mathrm{CO}}=40P^{-2}\mathrm{exp}(25,000/T)$~s, $\tau_{\mathrm{CO}_2}=10^{-10}P^{-0.5}\mathrm{exp}(38,000/T)$~s, $\tau_{\mathrm{NH}_3}=10^{-7}P^{-1}\mathrm{exp}(52,000/T)$~s, and $\tau_{\mathrm{CH}_4}=3\times 10^{-6}P^{-1}\mathrm{exp}(42,000/T)$~s. 
Here, $T$ and $P$ are the air temperature and pressure. The chemical timescales reported in \citet{zahnle_methane_2014} are based on solar composition, and variations in metallicity can influence them. For example, \citet{visscher_chemical_2012} suggested that $\tau_{\mathrm{CO}}$ scales with the square of metallicity, while $\tau_{\mathrm{CH}_4}$ is inversely proportional to metallicity. 
However, on the temperate mini-Neptune K2-18b, the dominant factor that increases $\tau_\mathrm{chem}$ is the low air temperature, while the impact of metallicity remains relatively minor. 
The chemical timescale of H$_2$O is not provided in \citet{zahnle_methane_2014}, but we expect its trend to agree with the four chemical species shown in the plot.
$\tau_\mathrm{mix}$ and $\tau_\mathrm{adv}$ are calculated based on the simulation results of the synchronous fixed abundance run.}
\label{fig:tscale}
\end{figure}

Fig.~\ref{fig:tscale} compares $\tau_\mathrm{chem}$ of CO, CO$_2$, NH$_3$, and CH$_4$ with $\tau_\mathrm{mix}$ and $\tau_\mathrm{adv}$ in the synchronous fixed-abundance run. $\tau_\mathrm{chem}$ is calculated using the Arrhenius-like fits provided in \citet{zahnle_methane_2014}. 
$\tau_\mathrm{mix}$ is estimated as $H/w_\mathrm{rms}$, where $H$ is the atmospheric scale height and $w_\mathrm{rms}$ is the horizontal root-mean-square of vertical velocity. $\tau_\mathrm{adv}$ is given by $R_\mathrm{p}/u_\mathrm{rms}$, where $R_\mathrm{p}$ is the planetary radius and $u_\mathrm{rms}$ is the horizontal root-mean-square of zonal wind speed. At pressure levels above approximately 10~bar, chemical timescales become significantly longer than both $\tau_\mathrm{mix}$ and $\tau_\mathrm{adv}$, indicating that atmospheric dynamics dominates in these regions. 

To diagnose the strength of vertical transport, characterize horizontal transport, and estimate equivalent 1D $K_{zz}$ profiles in regions dominated by atmospheric dynamics, we incorporate passive tracers into the kinetics runs. These passive tracers are advected solely as mass mixing ratios by the flow in the GCM. They do not affect radiative transfer (opacity) nor participate in chemical reactions. It is worth noting that tracer transport is computed concurrently with the chemical species and reactions. We focus on the pressure levels where atmospheric transport dominates, so no source/sink terms are applied to the tracer field:
\begin{equation}
    \frac{dq}{dt} = 0.
\end{equation}
Here, $q$ is the tracer mass mixing ratio and $\frac{dq}{dt}$ is the material derivation of the tracer mass mixing ratio. Thus, the tracers' local change over time is determined by advection. In the spherical coordinates, this yields:
\begin{equation}
\frac{\partial q}{\partial t} = -\frac{u}{r\mathrm{cos}\phi}\frac{\partial q}{\partial \lambda}-\frac{v}{r}\frac{\partial q}{\partial \phi} -w\frac{\partial q}{\partial r} .
\end{equation}
In the above equation, $u$, $v$, and $w$ are the zonal, meridional, and vertical velocities, and the coordinates $\lambda$, $\phi$, and $r$ are longitude, latitude, and radial distance. 

The initial tracer profile $q_\mathrm{init}$ is taken to be a simple function in which  $q$ exponentially decreases from a fixed abundance at the quench pressure $P_\mathrm{quench}$:
\begin{equation}
q_\mathrm{init} = 
\left\{
\begin{array}{ll}
    10^{-5}\cdot\left( \frac{P}{P_\mathrm{quench}} \right)^{1.5} &  P < P_\mathrm{quench} \\
    10^{-5} &  P \geq P_\mathrm{quench}
\end{array}
\right.
\end{equation}
We set $P_\mathrm{quench}=10$~bar as $\tau_\mathrm{chem}$ becomes significant larger than $\tau_\mathrm{mix}$ and $\tau_\mathrm{adv}$ at 10~bar. This setup represents a chemical species with an extremely long $\tau_\mathrm{chem}$, sourced uniformly from the deep atmosphere.
The investigation of tracers sourced from the deep atmosphere is motivated by the importance of CO$_2$ quenching in explaining the transmission spectra reported in \citet{madhusudhan_carbon-bearing_2023}, as suggested by \citet{wogan2024jwst}.
The choice of the quench pressure is supported by the fact that the quenched levels for most of the dominated species range between 1 and 10~bar in 3D simulations, as discussed in Part II.  
The analysis and the estimated equivalent 1D $K_{zz}$ profile are insensitive to the initial tracer mass mixing ratio, provided that the initial vertical gradient of the tracer is sufficiently steep and the simulations run long enough to achieve a quasi-steady state \citep{komacek_vertical_2019}. 

Note that previous studies on vertical mixing in exoplanetary atmospheres typically include source and sink terms for tracers representing chemical species \citep[e.g.,][]{zhang_global-mean_2018, Zhang_2018_II, komacek_vertical_2019}. This is primarily because those studies focus on hotter planets, such as hot Jupiters, which exhibit relatively short chemical timescales. For instance, the longest $\tau_\mathrm{chem}$ considered in \citet{komacek_vertical_2019} and \citet{Zhang_2018_II} are $10^6$~s and $10^{11}$~s, respectively, with the latter value occurring only in the uppermost atmosphere during their `quench experiments', where $\tau_\mathrm{chem}$ increases with altitude. 
In contrast, the $\tau_\mathrm{chem}$ we estimate using our temperature profiles exceeds 10$^7$~s above 10~bar and exceeds 10$^{19}$~s above 1~bar, indicating significantly slower chemical processes. Our study thus addresses tracer behavior with extremely long lifetimes--a regime not explored in previous works. Moreover, because our simulations include realistic chemical reactions, we are able to directly determine quench levels without requiring additional parameter studies \citep{Zhang_2018_II}.

\section{Results} \label{sec:results}

In this section, we present the results and analysis of our 3D chemical kinetics simulations for K2-18b. Sections~\ref{subsec:thermal} and~\ref{subsec:circulation} explore the thermal structure and atmospheric circulation, respectively, while section~\ref{subsec:tracer} discusses the three-dimensional distribution of passive tracers used to investigate atmospheric transport processes.

\subsection{Thermal structure} \label{subsec:thermal}

\begin{figure*}
\centering
\includegraphics[width=2\columnwidth]{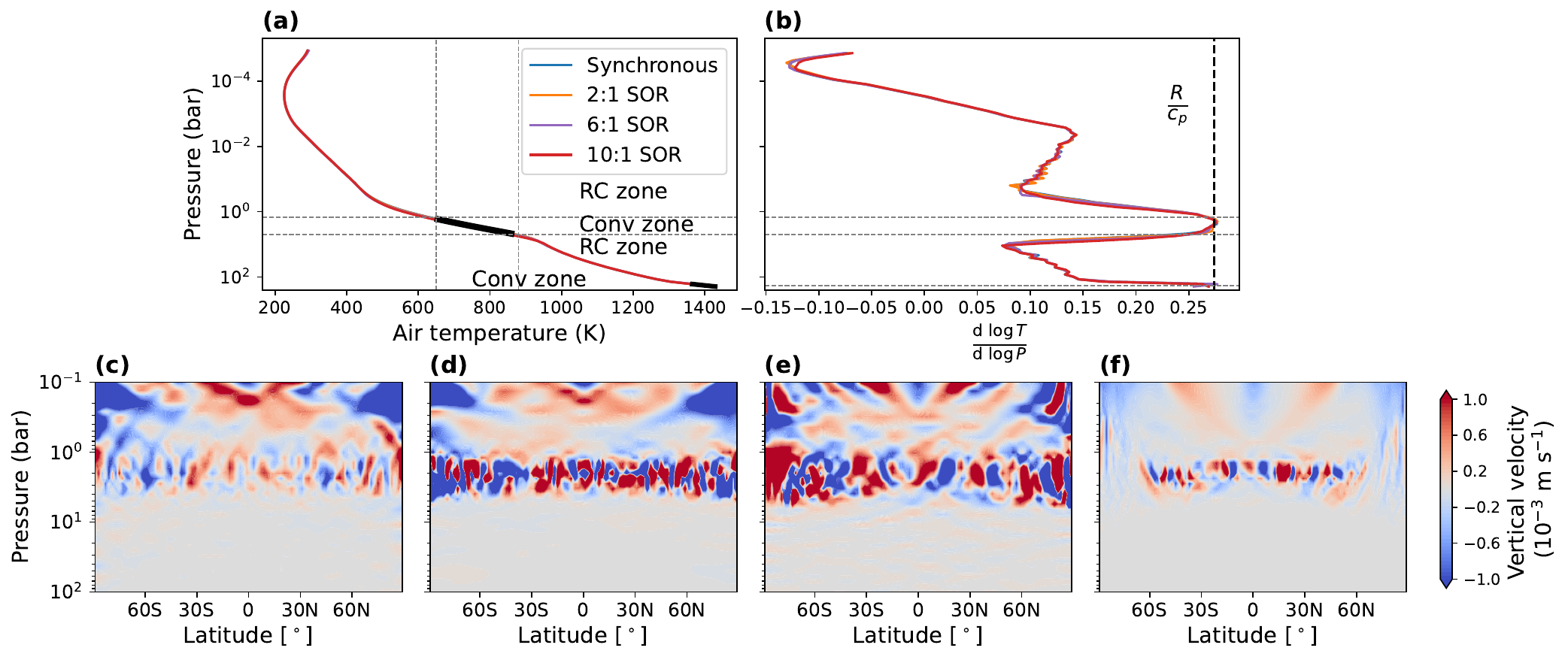}
\caption{Global mean air temperature vertical profiles (a), global mean air temperature lapse rate profiles (b), and zonal-mean vertical velocity between 0.1 and 200~bar ((c)--(f)). We restrict the vertical velocity plots to this pressure range to better highlight the enhanced vertical motion within the detached convective zone, located between 1 and 5~bar.
In panels (a) and (b), different coloured lines represent results from simulations with different rotation periods. Coloured lines in panel (a) overlap with each other, indicating that global mean air temperature profiles are consistent between different rotation periods.
The dashed black line in panel (b) is the adiabatic temperature lapse rate, and its intersection with the local temperature lapse rate indicates the appearance of convective zones. The thin gray dashed lines indicate the boundary between the radiative-convective (RC) zones and the convective (Conv) zone.}
\label{fig:tair_profile}
\end{figure*}

\begin{figure*}
\centering
\includegraphics[width=2\columnwidth]{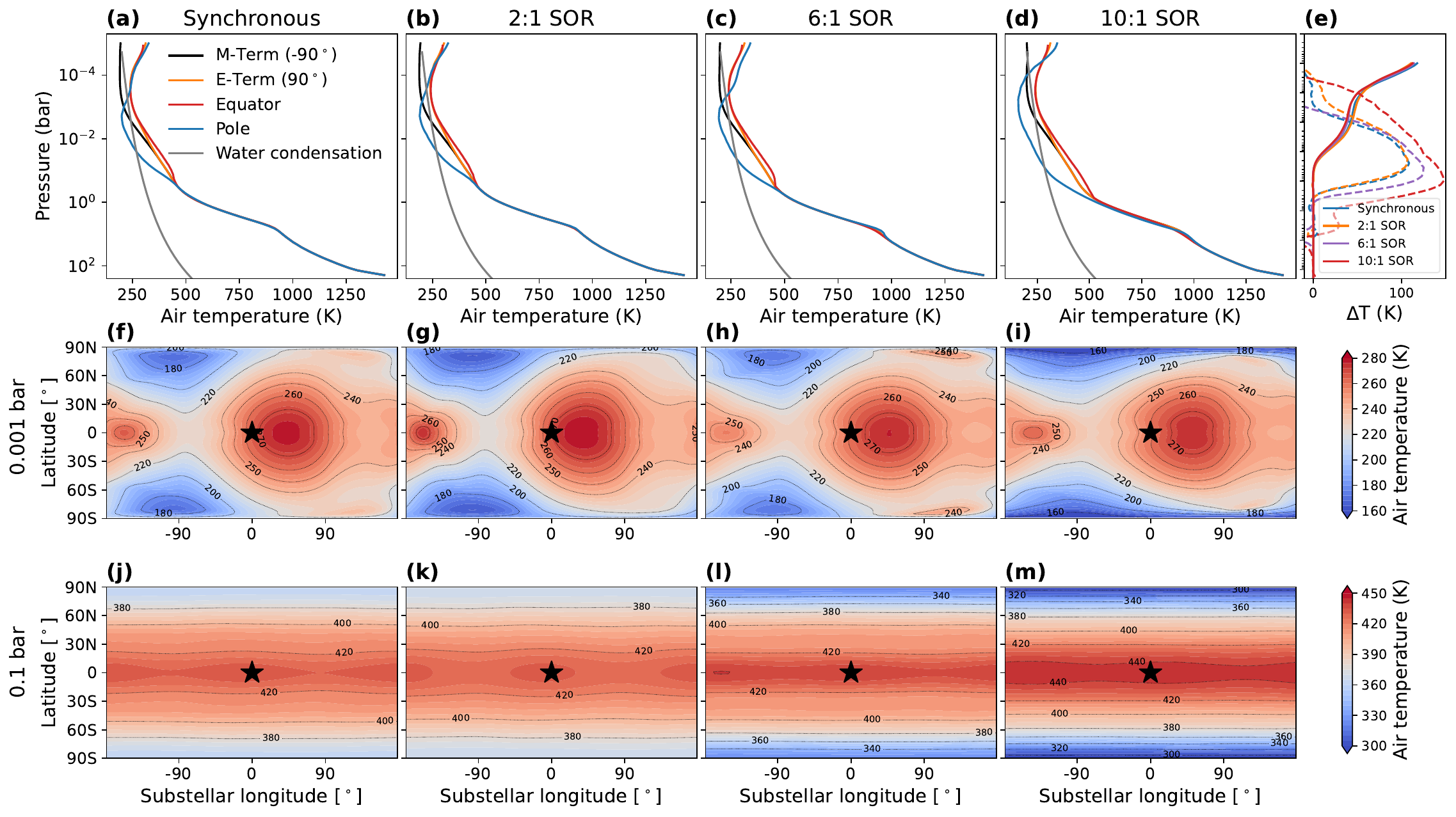}
\caption{Air temperature vertical profiles ((a)--(e)) and air temperature horizontal distribution at 0.001~bar ((f)--(i)) and 0.1~bar ((j)--(m)). The results are transforming in the heliocentric frame (keeping the substellar point at 0$^{\circ}$ longitude and 0$^{\circ}$ latitude). In panels (a)--(d), zonal-mean air temperatures at the equator (0$^\circ$) and the polar regions (average over 90$^\circ$S and 90$^\circ$N) are shown in red and blue lines, black and orange lines show the meridional-mean air temperature at the morning and evening terminators, and water condensation curves are shown in gray lines.
Panel (e) depicts the temperature contrast between the evening and morning terminators (solid) and the temperature contrast between the equator and the poles (dashed). Coloured lines in panel (e) show results from different simulations. The black star-shaped markers in panels (f)--(m) indicate the location of the substellar point.
The first four columns from left to right are results from simulations assuming K2-18b is synchronous or is under 2:1 SOR, 6:1 SOR, and 10:1 SOR. }
\label{fig:thermal_structure}
\end{figure*}

In this section, we analyse the thermal structure of K2-18b.
As shown in Fig.~\ref{fig:tair_profile}(a), the global mean air temperature vertical profiles of K2-18b are consistent between different rotation periods. For pressures larger than 0.0005~bar, air temperature increases with pressure, while in the upper atmosphere, a thermal inversion forms due to the strong shortwave absorption of CH$_4$, whose value mixing ratio is $\sim$0.1 at these layers (will be shown in Part II).

K2-18b's thermal structure features two detached convective zones, situated between radiative-convective regions. These convective zones occur at approximately 1 to 5~bar and in the deep atmosphere around 190~bar, as highlighted by the thick black line in Fig.~\ref{fig:tair_profile}(a).
These convective regions are characterized by the air temperature following the dry adiabatic lapse rate (Fig.~\ref{fig:tair_profile}(b)), where $\mathrm{d log}\,T/\mathrm{d log}\,p = R/c_p=0.274$. Here, $T$ and $p$ are air temperature and pressure, $R$ is the gas constant, and $c_p$ is the specific heat capacity. As mentioned in section~\ref{subsec:K2-18b set up}, $R$ and $c_p$ are set to constant values based on the vertical mean from the 1D \texttt{ATMO} chemical kinetics simulation. However, in reality, both parameters can vary with altitude (Fig.~\ref{Cp_R}).

The convective zones form because radiative processes at these layers are insufficient to carry energy out. The peak of the Planck function in the detached convective zone, between 1 to 5~bar with temperatures of 650 to 880~K, falls within the 4.5 to 3.3~$\mu$m range determined by Wien's Displacement Law ($\lambda_\mathrm{max} = 2.897\times10^{-3}/T$). This places the local Planck function directly within the strong CO$_2$ absorption band at 4.2~$\mu$m \citep{yurchenko2020exomol} and the strong CH$_4$ absorption band at 3.3~$\mu$m \citep{yurchenko_exomol_2024}. As will be shown in Part II, CO$_2$ and CH$_4$ are two dominant chemical species, with mole fractions of $\sim$10$^{-3}$ and $\sim$0.1 at this layer.
The strong absorption of CO$_2$ and CH$_4$ forms an optically thick layer, significantly reducing the thermal flux. Simultaneously, strong near-infrared absorption by CH$_4$ and H$_2$O limits the amount of stellar radiation reaching this region. With reduced radiative flux, heat transport is primarily driven by convection, further stimulating convective motions \citep{marley_cool_2015}. In deeper layers, the peak of the Planck function shifts to shorter wavelengths. At the deepest layer, where the temperature reaches approximately 1300~K, the Planck function peak moves to 2.3~$\mu$m, aligning with another strong CH$_4$ absorption band \citep{yurchenko_exomol_2024}, which leads to the formation of another convective layer. 

Notably, the detached convective region between 1 to 5~bar exhibits relatively strong vertical velocity (Figs~\ref{fig:tair_profile}(c)--(f)) and vertical mixing (discussed in section \ref{subsec:tracer}). The deep convective layer near the bottom boundary induces instability in the model, necessitating the implementation of a bottom sponge layer and bottom drag to maintain stability, as detailed in Appendix \ref{stability}.

To preserve the day-to-night thermal contrasts that might be smoothed out by temporal averaging, we present the horizontal thermal structures in a heliocentric frame, where the substellar point remains fixed at 0$^\circ$ longitude and 0$^\circ$ latitude. This approach is particularly important for asynchronous rotators, as the observed transmission spectrum probes the morning and evening terminators at the moment of transit. 

In the heliocentric frame, a temperature difference between the evening and morning terminators (referred to as the limb difference) emerges around 0.02~bar across all simulations, increasing from a few Kelvin at 0.02~bar to a maximum of 120~K at the top of the atmosphere (Figs~\ref{fig:thermal_structure}(a)--(e)). This limb difference remains largely consistent across different rotation rates (Fig.~\ref{fig:thermal_structure}(e)). It is primarily driven by eastward heat transport from the substellar point, caused by strong westerly winds in the upper atmosphere (discussed further in section~\ref{subsec:circulation}). A comparison of Figs~\ref{fig:thermal_structure}(f)--(i) shows that the eastward offset of the hotspot becomes more prominent in the 6:1 SOR and 10:1 SOR simulations, where faster rotation shortens the planetary day, thus reducing the substellar heating duration and enhancing the efficiency of zonal heat redistribution.

As shown in Fig.~\ref{fig:thermal_structure}(e), a noticeable meridional temperature contrast develops above 0.5~bar in the synchronous, 2:1 SOR, and 6:1 SOR simulations, while in the 10:1 SOR simulation, it emerges at a deeper level, above 10~bar.  
The meridional temperature contrast remains consistent in the synchronous and 2:1 SOR simulations but increases with rotation rate in the 6:1 SOR and 10:1 SOR simulations. This trend is more evident in Figs~\ref{fig:thermal_structure}(j)--(m), where the synchronous and 2:1 SOR simulations exhibit similar temperature patterns, both showing an equator-to-pole contrast of approximately $\sim$60~K. In contrast, the 6:1 SOR and 10:1 SOR simulations display a significantly larger meridional temperature difference, with the contrast increasing from $\sim$120~K in the 6:1 SOR simulation to $\sim$150~K in the 10:1 SOR simulation.

This trend can be explained by the dimensionless equatorial Rossby deformation length, given by
\begin{equation}
    \mathcal{L} = \frac{L_\mathrm{eq}}{R_\mathrm{p}}=\sqrt{\frac{NH}{2R_\mathrm{p}\Omega}}
\end{equation}
in which $L_\mathrm{eq}$ is the equatorial Rossby deformation radius, $N$ is the Brunt-Vaisala frequency ($N=\sqrt{\frac{g}{T}\left(\frac{g}{c_\mathrm{p}}+\frac{dT}{dr}\right)}$\,), where $g$ is surface gravity, $H$ is the atmospheric scale height ($\frac{RT}{g}$), and $\Omega$ is the rotation rate \citep{leconte20133d,showman_atmospheric_2020}. As seen in Fig.~\ref{fig:tair_profile}(a), the atmosphere is nearly isothermal at 0.001~bar ($|dT/dr|<$0.001~K\,m$^{-1}$), so $NH$ can be simplified to $R\sqrt{\frac{T}{c_p}}$. Use the parameters of K2-18b listed in Table \ref{k2-18b}, and using $T=$ 250~K at 0.001~bar, $\mathcal{L}$ for simulations of synchronous, 2:1 SOR, 6:1 SOR, and 10:1 SOR are 1.91, 1.35, 0.78, and 0.6, respectively. For simulations of synchronous and 2:1 SOR, $\mathcal{L}$ is larger than 1, indicating that the whole planet is predominantly in a dynamical  `tropical regime' \citep{andrews1987middle,holton2013introduction,Showman_circulation_2013}. In this state, due to the relatively weak effects of rotation, atmospheric waves (e.g., gravity waves) are prevalent at all longitudes, reducing the meridional temperature contrast. When the rotation period decreases to 5.49 days for simulation 6:1 SOR, $\mathcal{L}$ decreases below 1, suggesting that these tropical waves are trapped in lower latitudes and the temperature contrast increases. As the rotation period further decreases, $\mathcal{L}$ decreases, and the meridional temperature contrast continues to increase.

The water condensation curves are plotted in Figs~\ref{fig:thermal_structure}(a)--(d). The region enclosed by the intersection of condensation curves and air temperature profiles, between 0.01~bar and 10$^{-4}$~bar, indicates that if sufficient H$_2$O molecules and cloud condensation nuclei are present on K2-18b, H$_2$O can condense to form water clouds. These clouds may be populated at the polar regions and can form at the morning terminator in all the simulations. Although not included in the model, the moist-convection inhibition induced by H$_2$O condensation may alter the air temperature and reduce vertical mixing between 0.01~bar and 10$^{-4}$~bar \citep{guillot1996giant,Leconte_2017,Leconte_2024}.

\subsection{Atmospheric circulation}\label{subsec:circulation}

\begin{figure*}
    \centering
    \includegraphics[width=2\columnwidth]{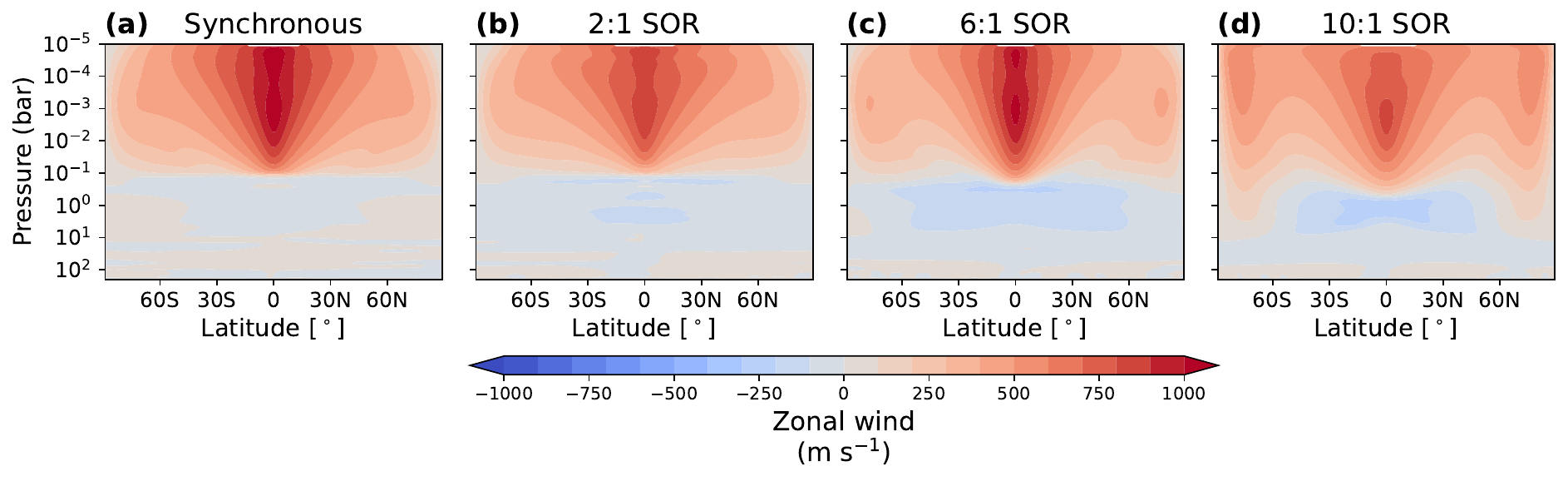}
    \caption{The zonal-mean zonal wind for all the simulations. }
    \label{fig:zonal_wind}
\end{figure*}

\begin{figure*}
\centering
\includegraphics[width=1.8\columnwidth]{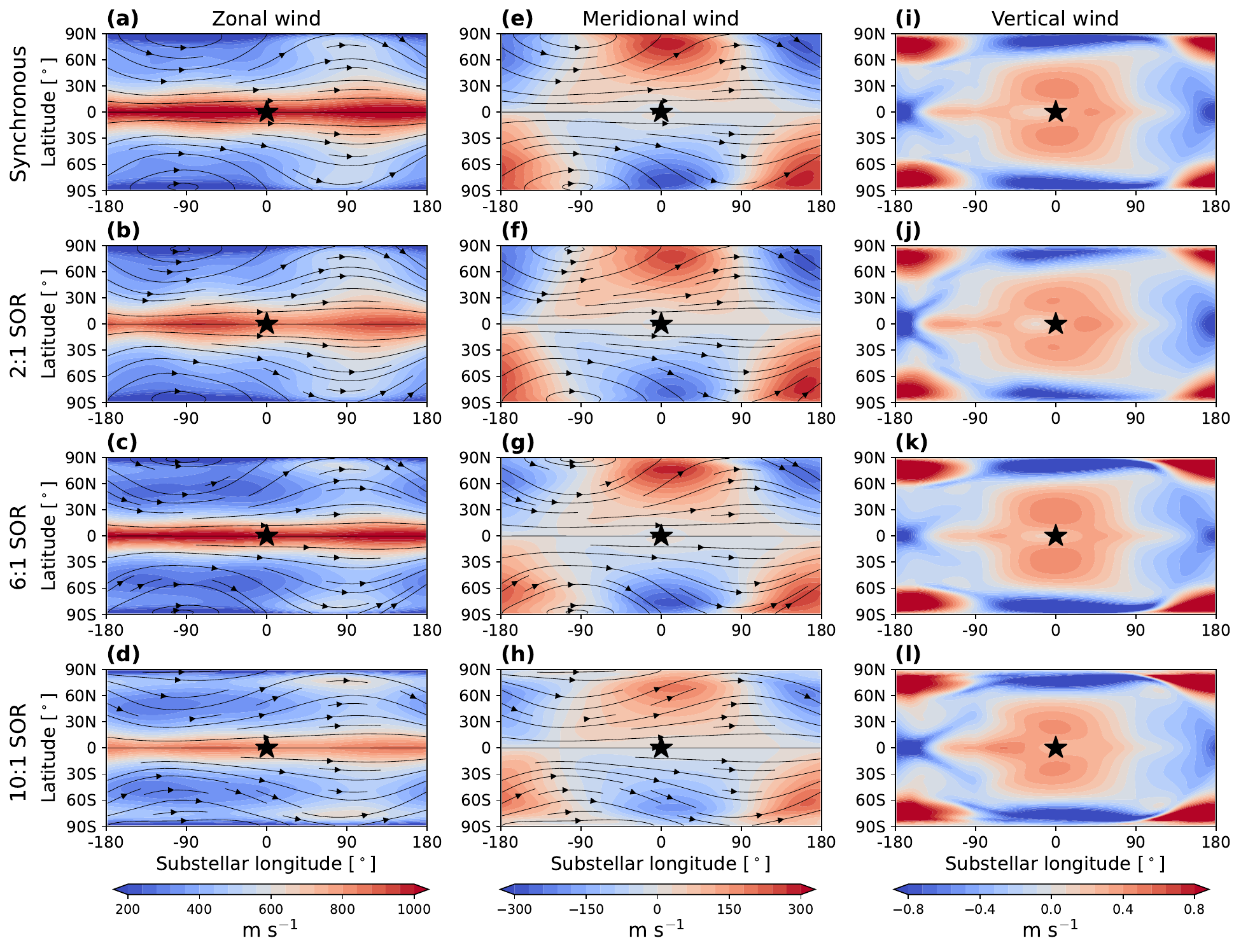}
\caption{Horizontal distribution of zonal wind ((a)--(d)), meridional wind ((e)--(h)), and vertical wind ((i)--(l)) at 0.001~bar. The streamlines indicate the direction of the horizontal flow. The results are transforming in the heliocentric frame to show the wind patterns more clearly. The black star-shaped markers indicate the location of the substellar point. Rows from the top to the bottom are results from simulations of synchronous, 2:1 SOR, 6:1 SOR, and 10:1 SOR.}
\label{fig:uvw}
\end{figure*}

\begin{figure*}
\centering
\includegraphics[width=2\columnwidth]{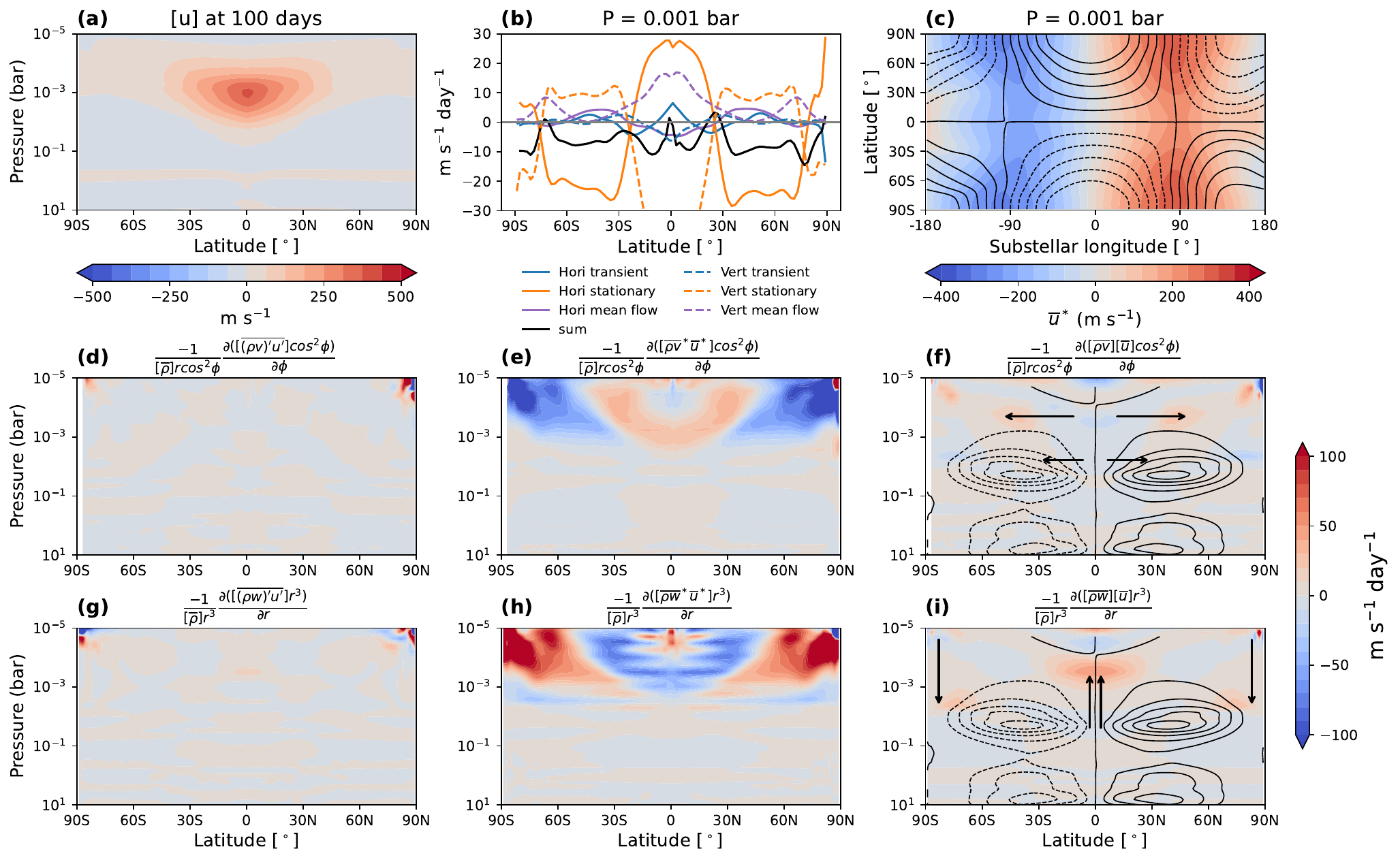}
\caption{Mechanism for the formation of the equatorial superrotating jet for the synchronous simulation. (a) The zonal-mean zonal wind at 100 days. (b) Zonal-mean zonal wind accelerations at 0.001~bar due to horizontal (solid) and vertical (dashed) transient eddy (blue), stationary eddy (orange), and mean flow (purple) in equation (\ref{momentum equation new}). The black line in panel (b) is the sum of these six terms. (c) Horizontal distribution of the stationary eddy velocity at 0.001~bar, where $\overline{u}^*$ is depicted as the coloured contour and $\overline{\rho v}^*/[\overline{\rho}]$ is shown as the contour lines, with an interval of 40 m\,s$^{-2}$ and limits of -200 and 200 m\,s$^{-2}$. (d)-(i): Contributions to zonal wind acceleration from horizontal transient eddies, horizontal stationary eddies, horizontal mean flow, vertical transient eddies, vertical stationary eddies, and vertical mean flow in equation (\ref{momentum equation new}). Contour lines in panels (f) and (i) show the zonal-mean mass streamfunction, with arrows indicating the direction of mean circulation and angular momentum transport by the mean flow. The contour lines have an interval of 10$^{10}$~kg\,s$^{-1}$, with limits of -5$\times$10$^{10}$ and 5$\times$10$^{10}$~kg\,s$^{-1}$. Results in panels (b)-(i) are derived from simulated data over the first 100 days, with an output interval of 10 days.}
    \label{fig:momentum_ini_t}
\end{figure*}

In this section, we analyse the wind structure in the simulations.
As shown in Fig.~\ref{fig:zonal_wind}, eastward wind dominates the pressure levels above 0.1~bar with the occurrence of an equatorial superrotating jet in all the simulations. For simulations assuming synchronous and 2:1 SOR, only the equatorial jet is present. As the rotation period decreases, two additional off-equator jets emerge at higher latitudes at pressures lower than $\sim$0.1~bar and $\sim$10~bar for 6:1 SOR and 10:1 SOR simulations, respectively (Figs~\ref{fig:zonal_wind}(c) and (d)). The off-equator jet speed is stronger in the faster rotating 10:1 SOR simulation. 

Fig.~\ref{fig:uvw} depicts the horizontal distribution of wind patterns at 0.001~bar in the heliocentric frame. The zonal wind is dominated by the eastward wind (Fig.~\ref{fig:uvw}(a)--(d)).
The meridional wind is characterized by a robust equator-to-pole circulation (Figs~\ref{fig:uvw}(e)--(h)) on the day side and a reversed circulation on the night side, which should be triggered by the day-side circulation. The peak speed of the meridional wind can reach up to 300 m\,s$^{-1}$ at 0.001~bar. The vertical velocity field aligns with these circulation patterns (Figs~\ref{fig:uvw}(i)--(l)). Upwelling motions dominate the low- and mid-latitudes of the day side, and downwelling motion dominates high latitudes. The night side exhibits a reversed circulation, with upwelling at high latitudes and downwelling at low- and mid-latitudes, which the day-side circulation should trigger to maintain mass conservation. These meridional and vertical wind patterns persist to 0.01~bar, but with a small wind speed (not shown).

Given that the eastward winds in the upper atmosphere play a crucial role in shaping the horizontal distribution of tracers, we provide an analysis of the mechanisms driving the equatorial superrotation.
The formation of the equatorial superrotating jet and high-latitude jets can be explained by analysing the zonal- and temporal-mean zonal momentum equation:
\begin{align}\label{momentum equation}
\frac{\partial([\overline{\rho}][\overline{u}])}{\partial t}  = 
& - \frac{1}{r \cos^2 \phi} \frac{\partial ([\overline{\rho v}][\overline{u}] \cos^2 \phi)}{\partial \phi} - \frac{1}{r^3}\frac{\partial ([\overline{\rho w}] [\overline{u}]r^3)}{\partial r}   \nonumber \\
& - \frac{1}{r \cos^2 \phi} \frac{\partial ([\overline{(\rho v)'u'} \cos^2 \phi])}{\partial \phi} - \frac{1}{r^3}\frac{\partial ([\overline{(\rho w)'u'}]r^3)}{\partial r} \nonumber \\
& - \frac{1}{r \cos^2 \phi} \frac{\partial ({[\overline{\rho v}^*\overline{u}^* }\cos^2 \phi])}{\partial \phi} - \frac{1}{r^3}\frac{\partial {([\overline{\rho w}^*\overline{u}^* }]r^3)}{\partial r}  \nonumber \\
& + 2\Omega \mathrm{sin}\phi [\overline{\rho v}] - 2\Omega \mathrm{cos}\phi [\overline{\rho w}] + [\overline{\rho G_\lambda}] \nonumber \\
& -\frac{\partial([\overline{\rho ' u'}])}{\partial t} - \frac{\partial([\overline{\rho}^*\overline{u}^*])}{\partial t}.
\end{align}
Here, overbars (primes) and brackets (asterisks) denote temporal and zonal averages (deviations), respectively. The first six terms on the right-hand side represent horizontal and vertical mean flow, transient eddy, and stationary eddy. The seventh and eighth terms correspond to the Coriolis forces, while the ninth term $G_\mathrm{\lambda}$ accounts for the body force applied in the zonal direction. The last two terms describe the temporal change of the perturbations in zonal momentum. 
Compared to the momentum equation in pressure coordinates \citep[][equation (28)]{showman_equatorial_2011}, the equation presented here is formulated in a non-hydrostatic height coordinate and incorporates an additional temporal deviation from the mean flow. A detailed derivation of equation (\ref{momentum equation}) is provided in Appendix \ref{derivation}.

\begin{figure*}
\centering
\includegraphics[width=2\columnwidth]{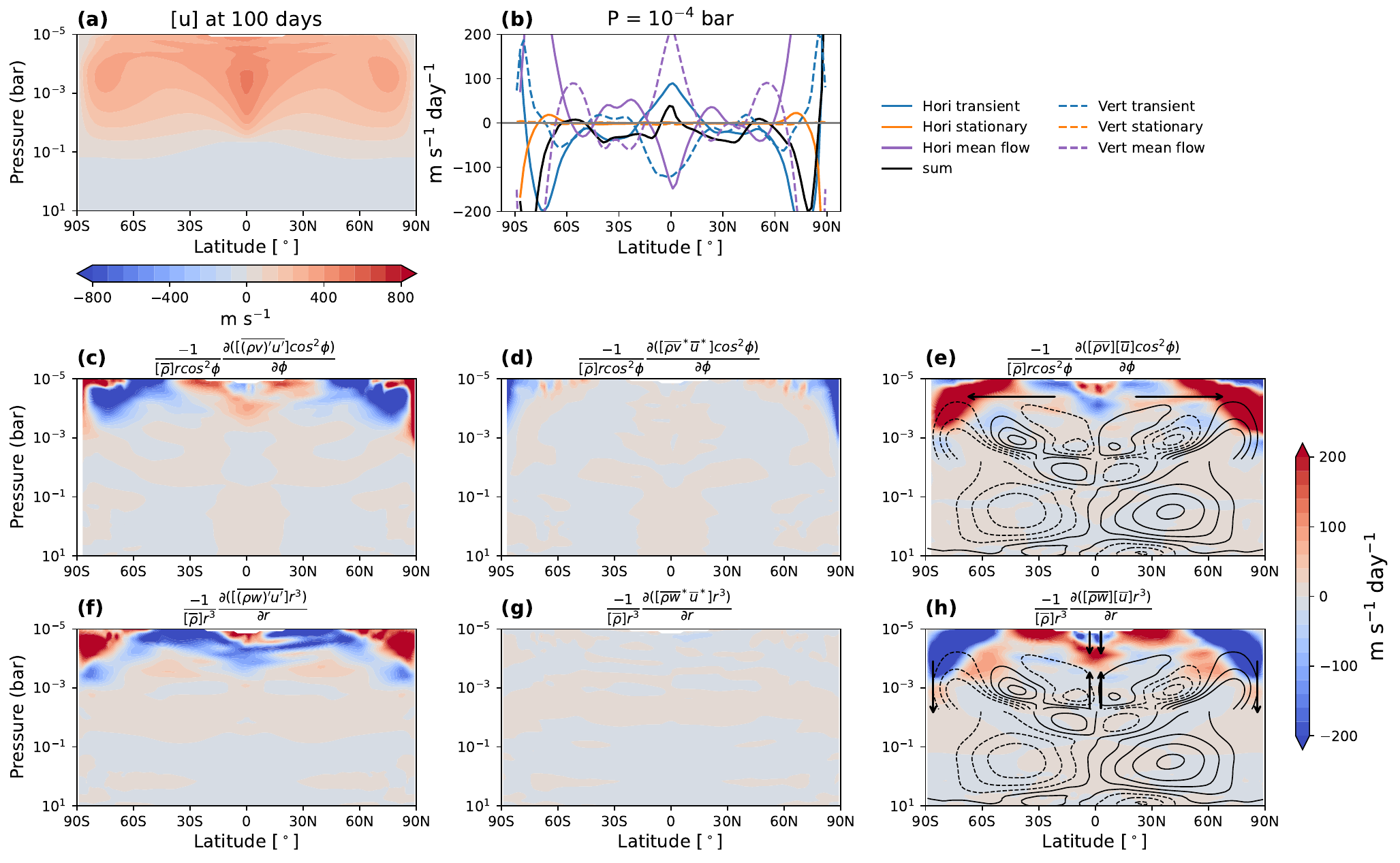}
\caption{Mechanisms of the formation of the equatorial superrotating jet and high-latitude jets for the 10:1 SOR simulation. (a) Zonal-mean zonal wind at 100 days. (b) Zonal-mean zonal wind accelerations at $10^{-4}$~bar due to horizontal (solid) and vertical (dashed) transient eddies (blue), stationary eddies (orange), and mean flow (purple) in equation (\ref{momentum equation new}). The black line in panel (a) represents the sum of these six terms. (c)-(h): Contributions to zonal wind acceleration from horizontal transient eddies, horizontal stationary eddies, horizontal mean flow, vertical transient eddies, vertical stationary eddies, and vertical mean flow in equation (\ref{momentum equation new}). Contour lines in panels (e) and (h) depict the zonal-mean mass streamfunction, with arrows indicating the direction of mean circulation and angular momentum transport by the mean flow. To enhance visibility of the mass streamfunction in the upper atmosphere, two contour intervals are applied: for pressures smaller than 0.006~bar, the interval is $8 \times 10^8$~kg\,s$^{-1}$ with limits of $-2 \times 10^9$ and $2 \times 10^9$~kg\,s$^{-1}$, while for pressures larger than 0.006~bar, the interval is $4 \times 10^9$~kg\,s$^{-1}$ with limits of $-2 \times 10^{10}$ and $2 \times 10^{10}$~kg\,s$^{-1}$. Results in panels (b)-(h) are derived from simulated data over the first 100 days, with an output interval of 10 days.}
    \label{fig:momentum_ini_10to1}
\end{figure*}

Considering the relatively smaller temporal change of $[\overline{\rho}]$ compared to $[\overline{u}]$ in the superrotation formation phase and to align the above equation to the zonal wind, we divide the equation (\ref{momentum equation}) by $[\overline{\rho}]$, and use this new momentum equation in the following analysis:
\begin{align}\label{momentum equation new}
\frac{\partial[\overline{u}]}{\partial t} \approx 
& - \frac{1}{[\overline{\rho}]r \cos^2 \phi} \frac{\partial ([\overline{\rho v}][\overline{u}] \cos^2 \phi)}{\partial \phi} - \frac{1}{[\overline{\rho}]r^3}\frac{\partial ([\overline{\rho w}] [\overline{u}]r^3)}{\partial r}  \nonumber\\
& - \frac{1}{[\overline{\rho}]r \cos^2 \phi} \frac{\partial ([\overline{(\rho v)'u'} \cos^2 \phi])}{\partial \phi} - \frac{1}{[\overline{\rho}]r^3}\frac{\partial ([\overline{(\rho w)'u'}]r^3)}{\partial r}  \nonumber\\
& - \frac{1}{[\overline{\rho}]r \cos^2 \phi} \frac{\partial ({[\overline{\rho v}^*\overline{u}^* }\cos^2 \phi])}{\partial \phi} - \frac{1}{[\overline{\rho}]r^3}\frac{\partial {([\overline{\rho w}^*\overline{u}^* }]r^3)}{\partial r} \nonumber \\
& + \frac{1}{[\overline{\rho}]} \{ 2\Omega \mathrm{sin}\phi[\overline{\rho v}] - 2\Omega \mathrm{cos}\phi [\overline{\rho w}] + [\overline{\rho G_\lambda}] \nonumber\\
&-\frac{\partial([\overline{\rho ' u'}])}{\partial t} - \frac{\partial([\overline{\rho}^*\overline{u}^*])}{\partial t}\}.  
\end{align}
Figs~\ref{fig:momentum_ini_t} and \ref{fig:momentum_ini_10to1} show the zonal- and temporal-mean momentum budget during the formation stage of the equatorial superrotating jets, using simulations of synchronous rotation and 10:1 SOR as examples. The meridional Coriolis term ($2\Omega\sin\phi[\overline{\rho v}]/[\overline{\rho}]$) is nearly zero at the equator and therefore does not contribute significantly to the formation of equatorial superrotation. The vertical Coriolis term ($2\Omega\cos\phi[\overline{\rho w}]/[\overline{\rho}]$) and body force ($[\overline{\rho G_\lambda}]/[\overline{\rho}]$) are small as the vertical wind is typically two to three orders of magnitude weaker than the horizontal winds and no drag is applied in the upper atmosphere.
The temporal changes of the perturbation  ($\frac{1}{[\overline{\rho}]}\frac{\partial([\overline{\rho'u'}])}{\partial t}$ and $\frac{1}{[\overline{\rho}]}\frac{\partial([\overline{\rho^*u^*}])}{\partial t}$) are negligible compared to the temporal change of the zonal wind ($\frac{\partial[\overline{u}]}{\partial t}$), as the density perturbations are two to three orders of magnitude smaller than the mean density. As a result, we focus on the first six terms of equation (\ref{momentum equation new}) in the analysis.

The superrotating jet forms within the first 100 days for the synchronous simulation (Fig.~\ref{fig:momentum_ini_t}(a)). The momentum budget at 0.001~bar, where the superrotation is the strongest, shows that the superrotation is mainly formed by horizontal stationary eddy convergence (Fig.~\ref{fig:momentum_ini_t}(b)). These stationary eddies are from the coupled Kelvin and Rossby waves, exhibiting a $\overline{\rho v}^*\overline{u}^*$ northwest-southeast phase tilt in the northern hemisphere and southwest-eastward phase tilt in the southern hemisphere at the west of the substellar point (Fig.~\ref{fig:momentum_ini_t}(c)), pumping the momentum to the equator \citep{showman_equatorial_2011}. To the east of the substellar point, reverse phase tilts occur, which partially cancel the acceleration. However, this cancellation is less significant, so overall, the horizontal stationary eddy continues to converge momentum toward the equator (Fig.~\ref{fig:momentum_ini_t}(b)). Above $\sim$0.001~bar, angular momentum is transported from the high latitudes to the lower latitudes due to the horizontal stationary eddy  (Fig.~\ref{fig:momentum_ini_t}(e)). The vertical stationary eddy convergence pattern at these levels is almost opposite to that of the horizontal stationary eddy convergence, showing a convergence at high latitudes and divergence at low latitudes  (Fig.~\ref{fig:momentum_ini_t}(h)). This suggests that the vertical stationary eddy is balancing the horizontal stationary eddy and vertically redistributing the momentum convergence. 

In the synchronous simulation, horizontal transient eddy also contributes to the formation of the superrotating jet at 0.001~bar, while compared to the horizontal stationary eddy, its strength is smaller (Figs~\ref{fig:momentum_ini_t}(d) and (e)). The vertical transient eddy is also weaker than the vertical stationary eddy (Figs~\ref{fig:momentum_ini_t}(g) and (h)). This is because, for the synchronous simulation with a fixed substellar point, zonal deviations are more significant than temporal deviations. 

The mean circulation acts to redistribute momentum in the synchronous simulation (illustrated by arrows). Above $\sim$0.01~bar, the equator-to-pole horizontal mean flow transports angular momentum from the low latitudes, where horizontal eddies accelerate zonal wind, to higher altitudes, where horizontal eddies decelerate zonal wind  (Fig.~\ref{fig:momentum_ini_t}(f)). This results in an angular momentum convergence at high latitudes and divergence at low latitudes by the mean horizontal circulation. At the equator, the upward mean flow transports angular momentum from the deeper layer where vertical eddies converge ($\sim$0.01~bar) to the upper atmosphere where vertical eddies diverge ($<$0.005~bar), accelerating the superrotating jet above 0.01~bar (Fig.~\ref{fig:momentum_ini_t}(i)). The downward mean flow near the poles transports angular momentum downward to $\sim$0.005~bar, accelerating the wind speed there.

In the simulation of 10:1 SOR, the equatorial superrotating jet and high-latitude jets become apparent features within the first 100 days of the simulation (Fig.~\ref{fig:momentum_ini_10to1}(a)). The momentum budget shows that at 10$^{-4}$~bar, superrotation is formed from horizontal transient eddy convergence and vertical mean flow (Fig.~\ref{fig:momentum_ini_10to1}(b)). Above 10$^{-4}$~bar, the horizontal transient eddy transports angular momentum from high latitudes to low latitudes, accelerating the equatorial superrotating jet (Fig.~\ref{fig:momentum_ini_10to1}(c)). Vertical transient eddy acts to cancel the acceleration of horizontal transient eddy at the equator by transporting angular momentum from low latitudes to high latitudes (Fig.~\ref{fig:momentum_ini_10to1}(f)). At 10$^{-4}$~bar, the equator-to-pole horizontal mean circulation takes away zonal momentum from the equator and redistributes it to higher latitudes (Fig.~\ref{fig:momentum_ini_10to1}(e)). This angular momentum loss at the equator is further compensated by the vertical transport of zonal momentum from the deeper and upper atmosphere to this level (Fig.~\ref{fig:momentum_ini_10to1}(h)), so the superrotation can be maintained and accelerated. 

The off-equatorial jets in the 10:1 SOR simulations are the combined result of horizontal mean circulation, vertical mean circulation, and horizontal transient eddy. As mentioned above, at 10$^{-4}$~bar, the horizontal mean flow transports zonal momentum from the equator to higher latitudes to accelerate the flows at high latitudes (Fig.~\ref{fig:momentum_ini_10to1}(e)). The downward mean flow can transport zonal momentum from upper levels ($>$0.001~bar) to $\sim$0.1~bar at high altitudes ($>$70$^\circ$), accelerating the wind there to make the high-latitude jets extend to the deeper layer (Fig.~\ref{fig:momentum_ini_10to1}(h)).
Moreover, the horizontal transient eddy convergence between $\sim$0.1 and $\sim$10~bar at latitudes higher than 70$^\circ$ further extends the high-latitude jets to deeper layers (Fig.~\ref{fig:momentum_ini_10to1}(c)). 

Different from the synchronous simulation, in the 10:1 SOR simulation, the transient eddies (Figs~\ref{fig:momentum_ini_10to1}(c) and (f)) are stronger than the stationary eddies (Figs~\ref{fig:momentum_ini_10to1}(d) and (g)), this is because as the substellar point moves along longitude through time, local temporal variations are more significant than the zonal perturbations.

\subsection{Passive tracer distribution} \label{subsec:tracer}

\begin{figure*}
\centering
\includegraphics[width=1.6\columnwidth]{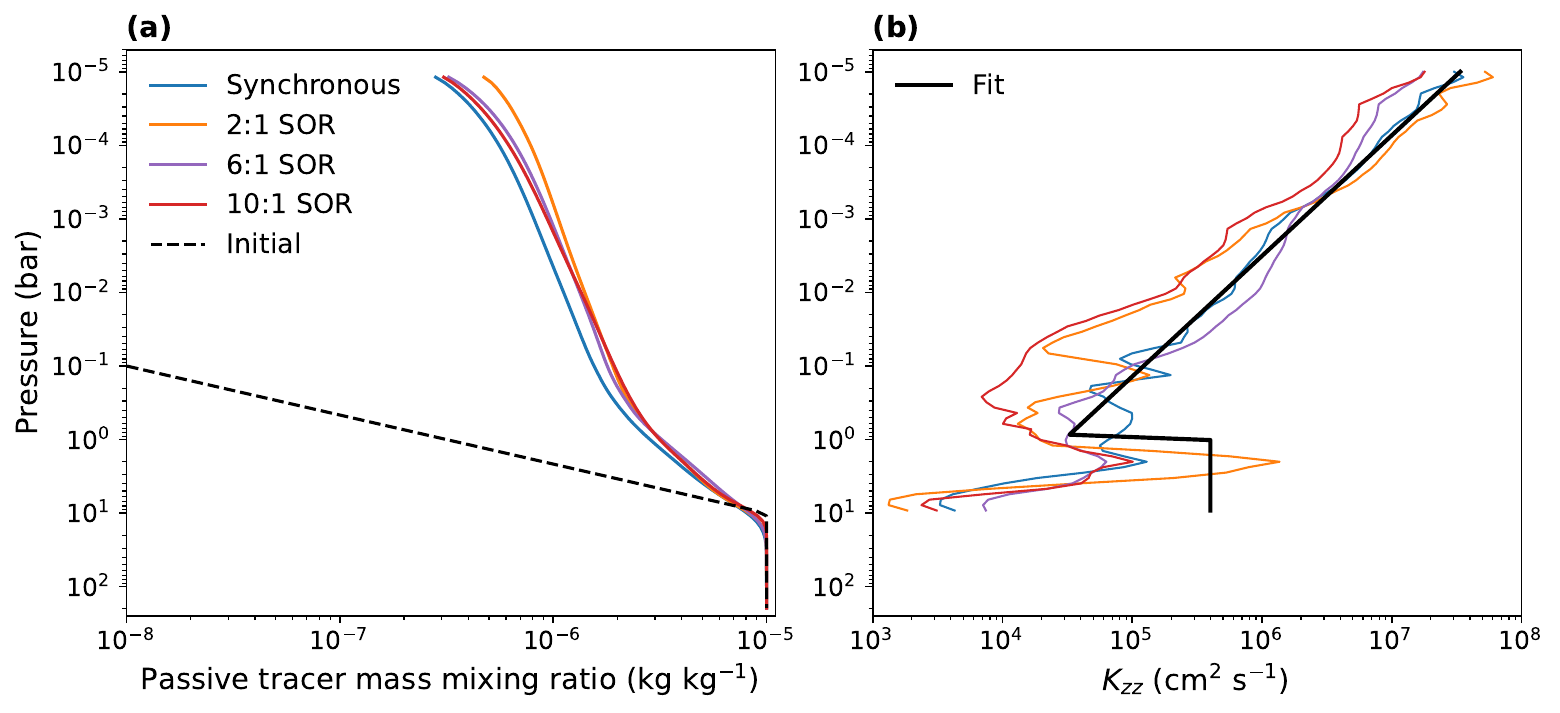}
\caption{Global mean vertical profiles of passive tracer mass mixing ratio (a) and equivalent 1D $K_{zz}$ profiles estimated from 3D simulations using the flux-gradient relationship (b). Different coloured lines represent the results of simulations with varying rotation periods. The dashed line in panel (a) indicates the initial passive tracer mass mixing ratio used in all simulations, and the thick black line in panel (b) suggests an `average' parameterization of the estimated $K_{zz}$ in different simulations. To obtain smoother $K_{zz}$ profiles, the results in panel (b) are averaged over the last 4600 days with an output interval of 10 days, after the model spin-up.}
\label{fig:tracer_gb}
\end{figure*}

\begin{figure*}
\centering
\includegraphics[width=1.6\columnwidth]{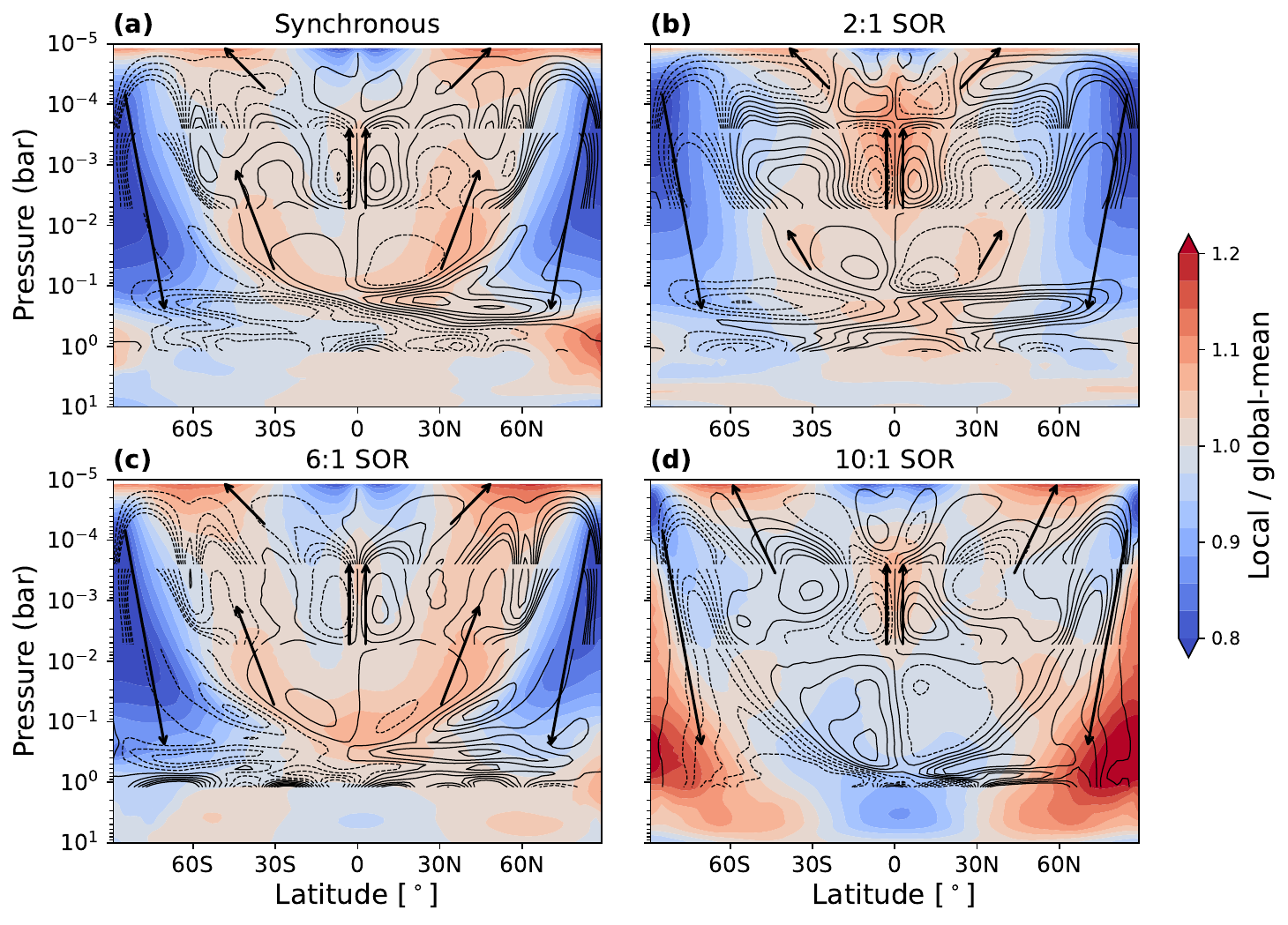}
\caption{Zonal-mean passive tracer mass mixing ratio (coloured contours) and zonal-mean mass streamfunction (contour lines) for different simulations. Coloured contours represent the ratio of local passive tracer mass mixing ratio to the global mean at a given isobar. Arrows indicate the direction of the zonal-mean circulation. To better visualize the mass streamfunction at different pressure levels, we apply three distinct contour ranges. For pressures larger than 0.006~bar, the contour interval is $4 \times 10^9$~kg\,s$^{-1}$, with limits of $-2 \times 10^{10}$ and $2 \times 10^{10}$~kg\,s$^{-1}$ for the synchronous, 2:1 SOR, and 6:1 SOR simulations; in the 10:1 SOR simulation, the contour interval is $8 \times 10^8$~kg\,s$^{-1}$, with limits of $-4 \times 10^9$ and $4 \times 10^9$~kg\,s$^{-1}$. For pressures ranging from 0.003 to 0.06~bar, the contour interval is $4 \times 10^8$~kg\,s$^{-1}$, with limits of $-2 \times 10^9$ and $2 \times 10^9$~kg\,s$^{-1}$ for all simulations. At pressures smaller than 0.003~bar, the contour interval is $8 \times 10^7$~kg\,s$^{-1}$, with limits of $-4 \times 10^8$ and  $4 \times 10^8$~kg\,s$^{-1}$ for all simulations.}
\label{fig:tracer_zonal}
\end{figure*}

In this section, we present the 3D tracer distribution and investigate how it is affected by the atmospheric circulation.
As shown in Fig.~\ref{fig:tracer_gb}, the global mean passive tracer mass mixing ratio is consistent across simulations with different rotation periods, indicating that rotation rate has little impact on the global mean vertical mixing strength. 

To analyse the strength of vertical mixing, we estimate the equivalent 1D $K_{zz}$ profile based on the flux-gradient relationship of the tracer fields in the 3D simulations using equation (23) in \citet {parmentier_3d_2013}:
\begin{equation}
\label{eq:flux-grandient}
    K_{zz} = -\frac{\langle\rho q w\rangle}{\langle\rho\frac{\partial q}{\partial r}\rangle}.
\end{equation}
This approach allows us to derive an equivalent vertical diffusive mixing rate from an inherently non-diffusive 3D atmosphere \citep{plumb1987zonally}. In the context of a 1D model, this vertical diffusive mixing rate is thought to reproduce the same vertical flux as that driven by dynamical processes in the 3D simulations. 
Note that this method can sometimes yield negative $K_{zz}$ values because the 3D atmosphere is not perfectly diffusive as the material surfaces of the extremely long-lived passive tracers are significantly distorted (i.e., horizontally non-uniformly distributed) \citep{zhang_global-mean_2018,Zhang_2018_II}. In such cases, we take the absolute value of equation (\ref{eq:flux-grandient}).
The 1D $K_{zz}$ profiles estimated from the 3D simulations are averaged over the last 4600 days after the model spin-up and are shown in Fig.~\ref{fig:tracer_gb}(b). These $K_{zz}$ profiles exhibit some dips and spikes due to vigorous local variations in the simulations. The $K_{zz}$ profiles generated from different simulations agree with each other in general, further supporting that the effect of the rotation rate on global mean vertical mixing is small.

The $K_{zz}$ profiles exhibit two distinct regions. At pressures smaller than 1~bar, $K_{zz}$ decreases exponentially with pressure, approximately following the parameterization:  $K_{zz} = 3\times10^4 \cdot\left(\frac{1}{P_{\mathrm{bar}}}\right)^{0.61}$~cm$^2$\,s$^{-1}$.
This behavior arises because strong vertical motions in these radiative-convective regions are primarily driven by the uneven distribution of received stellar flux. As pressure increases, radiative forcing weakens, and the atmospheric radiative timescale increases, leading to reduced vertical motions \citep[e.g.,][]{Iro_2005,Zhang_2018_II,komacek_vertical_2019}.  
At pressures between 1 and 5~bar, the presence of a detached convective zone induces more vigorous vertical motions, enhances vertical mixing, and consequently increases $K_{zz}$ ($\sim$$4\times10^5$~cm$^2\,$s$^{-1}$). A more detailed discussion of the estimated $K_{zz}$ profile will be presented in Part II.

Rotation rates significantly influence the meridional distribution of passive tracers. In the synchronous, 2:1 SOR, and 6:1 SOR simulations, the zonal-mean local tracer mass mixing ratio shows a $\sim$20$\%$ deviation from the global mean at each pressure level (Figs~\ref{fig:tracer_zonal}(a)--(c)). In contrast, the 10:1 SOR simulation exhibits a larger perturbation of $\sim$40$\%$ (Fig.~\ref{fig:tracer_zonal}(d)).

\begin{figure*}
\centering
\includegraphics[width=1.6\columnwidth]{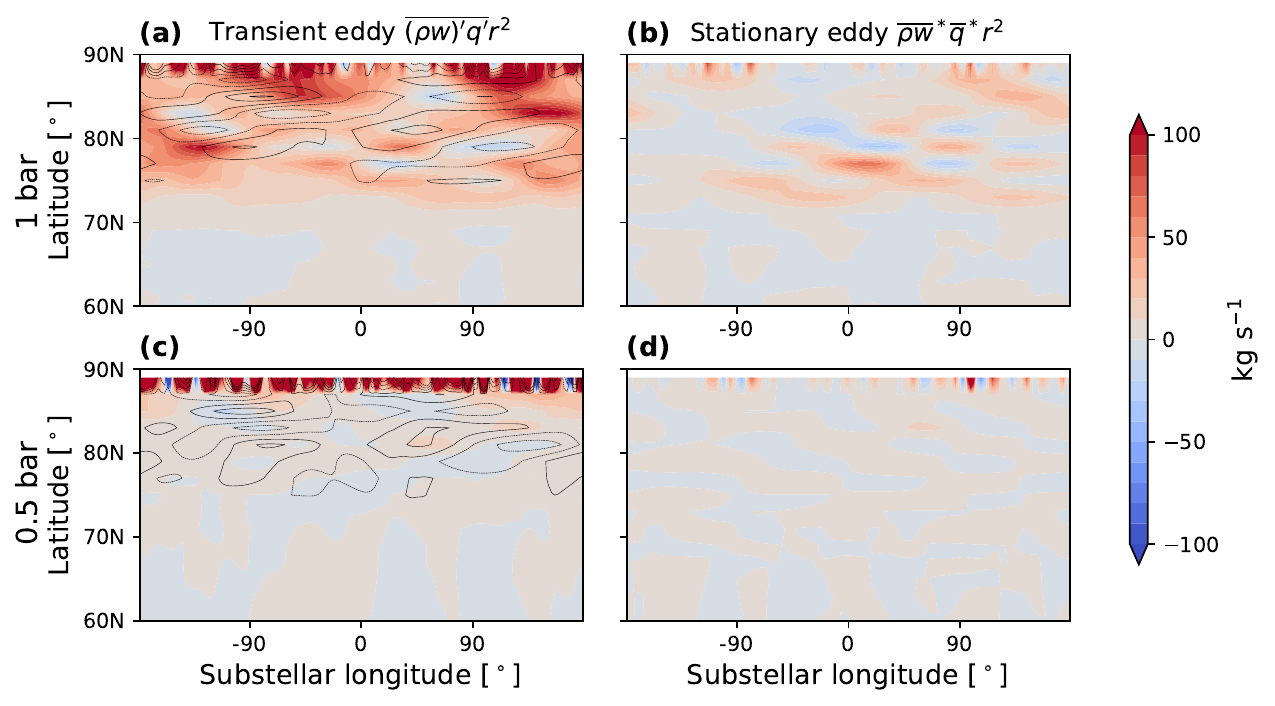}
\caption{Coluored contours show the horizontal distribution of vertical transient eddies ((a) and (c)) and stationary eddies ((b) and (d)) at 1~bar and 0.5~bar for the 10:1 SOR simulation. Contour lines in panels (a) and (b) represent vertical wind anomalies relative to the 330-day mean, evaluated on the final day of the simulation.}
\label{fig:tracer_eddy}
\end{figure*}

In the synchronous, 2:1 SOR, and 6:1 SOR simulations, tracers are concentrated at latitudes below 60$^\circ$ for pressure levels greater than 0.1~bar (Figs~\ref{fig:tracer_zonal}(a)--(c)). The tracer distribution closely follows the zonal-mean mass streamfunction: regions of high tracer mass mixing ratio coincide with areas of upward motion, while low tracer mass mixing ratios align with descending branches, demonstrating a positive correlation between tracer mass mixing ratio and vertical velocity \citep{holton_dynamically_1986,zhang_global-mean_2018,Zhang_2018_II}. This positive correlation arises because, in the upwelling region, tracers with higher mass mixing ratios are transported upward from their deep atmospheric source, whereas in the downwelling region, tracers with lower mass mixing ratios are brought downward from the upper atmosphere. Consequently, the tracer mass mixing ratio is increased in the upwelling region and reduced in the downwelling region along an isobar.

In the 10:1 SOR simulation, tracer mass mixing ratio remains positively correlated with vertical velocity at latitudes below 60$^\circ$ (Fig.~\ref{fig:tracer_zonal}(d)). However, at higher latitudes ($>$70$^\circ$), particularly between 0.1 and 1~bar, tracer-abundant regions coincide with the descending branches of the circulation, indicating a breakdown of this positive correlation at these pressures.

The more abundant tracers at higher latitudes ($>$70$^\circ$) between 0.1~bar and 1~bar in the simulation of 10:1 SOR are caused by the strong vertical transient eddy in these regions, which transports tracers from the deep atmosphere. The mean and eddy interaction equation for the zonal- and temporal-mean tracer mass mixing ratio in the spherical coordinates can be written as:
\begin{align}
\frac{\partial{([\overline{\rho}]}[\overline{q}])}{\partial{t}} = 
&-\frac{1}{r\cos\phi}\{\frac{\partial([\overline{\rho v}][\overline{q}]\cos\phi)}{\partial \phi}+\frac{\partial({[\overline{(\rho v)'q'}]\cos\phi)}}{\partial \phi}\nonumber\\
&+\frac{\partial([\overline{\rho v}^*\overline{q}^*]\cos\phi)}{\partial \phi}\}  -\frac{1}{r^2}\{\frac{\partial([\overline{\rho w}][\overline{q}]r^2)}{\partial r}\nonumber\\
& +\frac{\partial([\overline{(\rho w)'q'}]r^2)}{\partial r}+\frac{\partial([\overline{\rho w}^*\overline{q}^*]r^2)}{\partial r}\}  \nonumber \\
& - \frac{\partial([\overline{\rho'q'}])}{\partial t} -\frac{\partial([\overline{\rho}^*\overline{q}^*])}{\partial t}. 
\end{align}
The first six terms on the right-hand side are the increases in passive tracer mass mixing ratio due to the transport by horizontal (and vertical) mean flow, transient eddy, and stationary eddy, and the last two terms are the temporal change in temporal and zonal perturbations of tracer mass mixing ratio. Among them, vertical terms should be the most important for tracer vertical transport. The derivation of this equation can be found in Appendix \ref{derivation_tracer}.
As shown in Fig.~\ref{fig:tracer_eddy}(a), the 10:1 SOR simulation exhibits a strong vertical transient eddy, which aligns with the large vertical wind temporal anomalies at latitudes higher than 70$^\circ$ at 1~bar. As the strength of the eddy decreases with height, $\partial(\overline{(\rho w)'q'}r^2)/\partial r<0$ (Fig.~\ref{fig:tracer_eddy}(c)), this leads to more abundant tracers at these regions. Compared with Figs~\ref{fig:tracer_eddy}(a) and (b), the transient eddy is stronger than the stationary eddy, because the temporal perturbation is more significant than the zonal perturbation for the asynchronous rotator.

\begin{figure*}
\centering
\includegraphics[width=2\columnwidth]{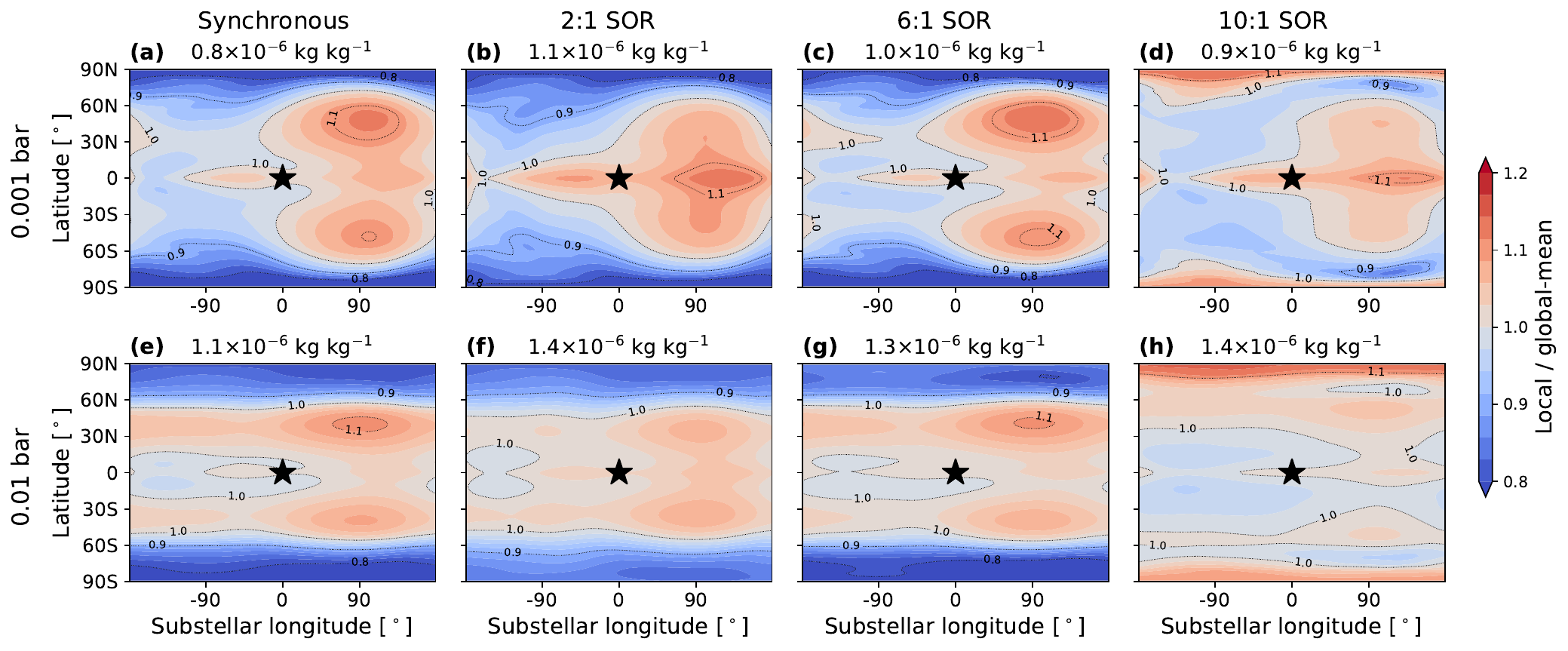}
\caption{The horizontal distribution of passive tracers at 0.001~bar ((a)--(d)) and 0.01~bar ((e)--(h)). The results are transforming into the heliocentric frame. The black star-shaped markers indicate the location of the substellar point. Coloured contours depict the ratio of local passive tracer mass mixing ratio to the global mean mass mixing ratio at certain pressure levels. Global mean values are printed above the plots. Columns from left to right are results from simulations of synchronous, 2:1 SOR, 6:1 SOR, and 10:1 SOR.}
    \label{fig:tracer_2D}
\end{figure*}

Fig.~\ref{fig:tracer_2D} shows the horizontal distribution of passive tracer mass mixing ratio at 0.001~bar and 0.01~bar. 
Latitudinally, at 0.001~bar, tracers are more abundant between 70$^\circ$S and 70$^\circ$N for simulations of synchronous, 2:1 SOR, and 6:1 SOR; while for the simulation of 10:1 SOR, tracers are abundant both at poles and between 70$^\circ$S and 70$^\circ$N (Figs~\ref{fig:tracer_2D}(a)--(d)). At 0.01~bar, in simulations of synchronous, 2:1 SOR, and 6:1 SOR, tracers are more abundant between 60$^\circ$S and 60$^\circ$N, while for the simulation of 10:1 SOR, it is more abundant at high latitudes (Figs~\ref{fig:tracer_2D}(e)--(h)). 

The more abundant passive tracers between 70$^\circ$S and 70$^\circ$N at 0.001~bar in all the simulations are caused by the strong upwelling at low- and mid-latitudes at the day side, triggered by the non-uniform stellar radiation as discussed in section \ref{subsec:circulation} (Figs~\ref{fig:uvw}(i)--(l)). The more abundant passive tracers at low- and mid-latitudes at 0.01~bar for simulations of synchronous, 2:1 SOR, and 6:1 SOR are also because of the day-side upwelling at these latitudes. 
In the 10:1 SOR simulation, passive tracers are more abundant at the poles at 0.001~bar and high latitudes at 0.01~bar due to the upwelling transport of tracers by transient eddies at these regions, as mentioned above.

Longitudinally, passive tracers accumulate preferentially near the evening terminator at 0.001~bar, with the tracer mass mixing ratio differing by approximately 20\% between the evening and morning terminators (Figs~\ref{fig:tracer_2D}(a)--(d)). This asymmetry arises because tracers are first transported upward from the deeper atmosphere by strong dayside upwelling and subsequently advected eastward by strong zonal winds (Figs~\ref{fig:uvw}(a)--(d)). Additionally, the strong meridional winds at 0.001~bar (Figs~\ref{fig:uvw}(e)--(h)) further redistribute tracers poleward, resulting in tracer-abundant regions extending beyond the upwelling zones.

At 0.01~bar, tracers are distributed more uniformly in the zonal direction compared to their distribution at 0.001~bar (Figs~\ref{fig:tracer_2D}(e)--(h)). This is because vertical motions are weaker and exhibit less zonal contrast at this pressure level, while the zonal wind remains sufficiently strong to homogenize tracer abundances longitudinally (figure not shown). The extent of tracer-abundant regions is also reduced at 0.01~bar, due to the weaker meridional winds relative to those at 0.001~bar (figure not shown).
 
\section{Conclusions and discussion}\label{conclusions}

In this study, we characterize the atmospheric circulation on temperate gas-rich mini-Neptune K2-18b and its impact on atmospheric transport by diagnosing the distribution of a passive tracer. This passive tracer is advected by wind flow but does not participate in chemical reactions or affect opacity. Given the comparable age of the K2-18 system and K2-18b's tidal spin-down timescale, we investigate both synchronous and asynchronous rotation states for K2-18b, considering SOR of 2:1, 6:1, and 10:1 in the latter. We assume K2-18b has 180 times solar metallicity, which is chosen based on the best agreement between the 1D model and the JWST observations reported in \citet{madhusudhan_carbon-bearing_2023}.

We find that a detached convective zone forms at pressures between 1 and 5~bar due to the strong thermal absorption of CH$_4$ and CO$_2$ in all the simulations. This detached convective zone can trigger vigorous convection and induce stronger vertical mixing. The upper atmosphere of K2-18b is dominated by eastward wind, with an equatorial superrotating jet occurring in all the simulations. As the rotation rate increases, two off-equatorial jets form in simulations of 6:1 SOR and 10:1 SOR. These westerly jets make the evening terminator hotter than the morning terminator in the upper atmosphere.
The formation of equatorial jets is due to the transport of eddy momentum from the high latitudes to the equator and from the vertical mean circulation, redistributing momentum from the vertical eddy momentum convergence region to the divergence regions. The off-equatorial jets form from the combined effects of the mean circulation and eddy momentum convergence. 

Rotation periods have minimal effects on the global mean air temperature profiles and the overall strength of the global mean vertical mixing. However, they significantly affect the latitudinal tracer distribution. For the synchronous, 2:1 SOR, and 6:1 SOR simulations, tracers are more abundant in the upwelling branches at low- and mid-latitudes and exhibit a positive correlation between passive tracer mass mixing ratio and vertical velocity. However, for the 10:1 SOR, passive tracers can be more abundant at high latitudes, aligning with the downwelling branch of the large-scale circulation because the strong vertical eddies at high latitudes can transport them from the deep atmosphere. The latitudinal differences in passive tracer mass mixing ratio can be up to $\sim$20$\%$ in the synchronous, 2:1 SOR, and 6:1 SOR simulations and $\sim$40$\%$ in the 10:1 SOR simulation at an isobar.

Longitudinally, we find that passive tracers accumulate more at the evening terminator in the upper atmosphere due to the strong westerly winds transporting them eastward from the substellar point, where the upwelling is most intense. This results in a $\sim$20$\%$ difference in passive tracer mass mixing ratio between the evening and morning terminators at an isobar. Since the passive tracer in our simulation represents long-lived chemical species, this suggests that such species may preferentially accumulate at the evening terminator, potentially introducing a limb asymmetry.
In Part II of this study, we will explicitly analyse chemical specie distribution and compute the resulting limb differences in the transmission spectrum. 

The similarity in global-mean temperature profiles and vertical mixing strength across simulations with varying rotation periods indicates that rotation has a limited influence on the global-mean vertical structure. Given that widely used 1D models aim to represent such global-mean conditions, our results support the continued use of 1D models for potentially asynchronously rotating planets such as K2-18b. However, once the latitudinal distribution of chemical species is considered, rotation becomes a critical factor. This has broader implications for temperate sub-Neptunes and Jupiters with long orbital periods that may not be in synchronous rotation. For these planets, incorporating rotation-driven dynamics may be essential to accurately capture atmospheric composition and circulation.

This study provides a detailed investigation into the role of atmospheric dynamics; however, even in regions where dynamics dominate, chemical processes remain influential in shaping the distribution of atmospheric species. For example, temperature contrasts at the quench level can lead to non-uniform abundances of chemical species, which in turn impact the horizontal distribution of chemical species in the upper atmosphere. These contrasts may arise from the uneven absorption of stellar flux or the radiative effects of chemical species deeper in the atmosphere. Together, these highlight the importance of fully coupled models that involve atmospheric dynamics, radiative transfer, and chemistry. In Part II of this study, we will analyse chemical structures resulting from the simulations, along with their interaction with radiative processes.

\section*{Acknowledgements}

We thank the anonymous referee for comments that improved the quality of this paper. We would like to thank the helpful discussion with Eva-Maria Ahrer. We want to thank Nathan Mayne and Krisztian Kohary for providing access to and assistance in running the UM. This work used the Max Planck Society's Viper High-Performance Computing system. This work also used the DiRAC Complexity system, operated by the University of Leicester IT Services, which forms part of the STFC DiRAC HPC Facility and the University of Exeter Supercomputer ISCA. We note that the production runs used in this paper required about 2.3 million CPU hours. The use due to testing and analysis is about 0.5 million CPU hours.
J. Liu is partly supported by the Overseas Study Program for Graduate Students of the China Scholarship Council. J.Y. is supported by the National Science Foundation of China (NSFC) under grant no. 42275134.

\section*{Data Availability}

The simulation data used in this study are available from the corresponding author upon reasonable request.



\bibliographystyle{mnras}
\bibliography{K2-18b_PartI} 

\begin{thebibliography}{}
\makeatletter
\relax
\def\mn@urlcharsother{\let\do\@makeother \do\$\do\&\do\#\do\^\do\_\do\%\do\~}
\def\mn@doi{\begingroup\mn@urlcharsother \@ifnextchar [ {\mn@doi@} {\mn@doi@[]}}
\def\mn@doi@[#1]#2{\def\@tempa{#1}\ifx\@tempa\@empty \href {http://dx.doi.org/#2} {doi:#2}\else \href {http://dx.doi.org/#2} {#1}\fi \endgroup}
\def\mn@eprint#1#2{\mn@eprint@#1:#2::\@nil}
\def\mn@eprint@arXiv#1{\href {http://arxiv.org/abs/#1} {{\tt arXiv:#1}}}
\def\mn@eprint@dblp#1{\href {http://dblp.uni-trier.de/rec/bibtex/#1.xml} {dblp:#1}}
\def\mn@eprint@#1:#2:#3:#4\@nil{\def\@tempa {#1}\def\@tempb {#2}\def\@tempc {#3}\ifx \@tempc \@empty \let \@tempc \@tempb \let \@tempb \@tempa \fi \ifx \@tempb \@empty \def\@tempb {arXiv}\fi \@ifundefined {mn@eprint@\@tempb}{\@tempb:\@tempc}{\expandafter \expandafter \csname mn@eprint@\@tempb\endcsname \expandafter{\@tempc}}}

\bibitem[\protect\citeauthoryear{Ag{\'u}ndez, Venot, Iro, Selsis, Hersant, H{\'e}brard  \& Dobrijevic}{Ag{\'u}ndez et~al.}{2012}]{agundez2012impact}
Ag{\'u}ndez M.,  Venot O.,  Iro N.,  Selsis F.,  Hersant F.,  H{\'e}brard E.,   Dobrijevic M.,  2012, \mn@doi [A\&A] {https://doi.org/10.1051/0004-6361/201220365}, 548, A73

\bibitem[\protect\citeauthoryear{Ag{\'u}ndez, Parmentier, Venot, Hersant  \& Selsis}{Ag{\'u}ndez et~al.}{2014}]{agundez2014pseudo}
Ag{\'u}ndez M.,  Parmentier V.,  Venot O.,  Hersant F.,   Selsis F.,  2014, \mn@doi [A\&A] {https://doi.org/10.1051/0004-6361/201322895}, 564, A73

\bibitem[\protect\citeauthoryear{Amundsen et~al.,}{Amundsen et~al.}{2016}]{amundsen2016uk}
Amundsen D.~S.,  et~al., 2016, \mn@doi [A\&A] {10.1051/0004-6361/201629183}, 595, A36

\bibitem[\protect\citeauthoryear{Andrews, Leovy  \& Holton}{Andrews et~al.}{1987}]{andrews1987middle}
Andrews D.~G.,  Leovy C.~B.,   Holton J.~R.,  1987, Middle atmosphere dynamics.
Academic press, \mn@doi{https://doi.org/10.1002/qj.49711548612}

\bibitem[\protect\citeauthoryear{{Baeyens}, {Decin}, {Carone}, {Venot}, {Ag{\'u}ndez}  \& {Molli{\`e}re}}{{Baeyens} et~al.}{2021}]{Baeyens_2021}
{Baeyens} R.,  {Decin} L.,  {Carone} L.,  {Venot} O.,  {Ag{\'u}ndez} M.,   {Molli{\`e}re} P.,  2021, \mn@doi [\mnras] {10.1093/mnras/stab1310}, \href {https://ui.adsabs.harvard.edu/abs/2021MNRAS.505.5603B} {505, 5603}

\bibitem[\protect\citeauthoryear{{Barrier} \& {Madhusudhan}}{{Barrier} \& {Madhusudhan}}{2025}]{Barrier_convection_2025}
{Barrier} E. F.~L.,  {Madhusudhan} N.,  2025, \mn@doi [\mnras] {10.1093/mnras/staf267}, \href {https://ui.adsabs.harvard.edu/abs/2025MNRAS.538.2463B} {538, 2463}

\bibitem[\protect\citeauthoryear{Batalha}{Batalha}{2014}]{batalha2014exploring}
Batalha N.~M.,  2014, \mn@doi [PNAS] {https://doi.org/10.1073/pnas.1304196111}, 111, 12647

\bibitem[\protect\citeauthoryear{Benneke et~al.,}{Benneke et~al.}{2019}]{benneke_water_2019}
Benneke B.,  et~al., 2019, \mn@doi [ApJ] {10.3847/2041-8213/ab59dc}, 887, L14

\bibitem[\protect\citeauthoryear{Benneke et~al.,}{Benneke et~al.}{2024}]{benneke2024jwst}
Benneke B.,  et~al., 2024, \mn@doi [arXiv preprint arXiv:2403.03325] {10.48550/arXiv.2403.03325}

\bibitem[\protect\citeauthoryear{Blain, Charnay  \& Bézard}{Blain et~al.}{2021}]{blain_1d_2021}
Blain D.,  Charnay B.,   Bézard B.,  2021, \mn@doi [A\&A] {10.1051/0004-6361/202039072}, 646, A15

\bibitem[\protect\citeauthoryear{Brande et~al.,}{Brande et~al.}{2024}]{brande_clouds_2024}
Brande J.,  et~al., 2024, \mn@doi [ApJ] {10.3847/2041-8213/ad1b5c}, 961, L23

\bibitem[\protect\citeauthoryear{Bézard, Charnay  \& Blain}{Bézard et~al.}{2022}]{bezard_methane_2022}
Bézard B.,  Charnay B.,   Blain D.,  2022, \mn@doi [Nature Astron.] {10.1038/s41550-022-01678-z}, 6, 537

\bibitem[\protect\citeauthoryear{Charnay, Blain, B{\'e}zard, Leconte, Turbet  \& Falco}{Charnay et~al.}{2021}]{charnay2021formation}
Charnay B.,  Blain D.,  B{\'e}zard B.,  Leconte J.,  Turbet M.,   Falco A.,  2021, \mn@doi [A\&A] {10.1051/0004-6361/202039525}, 646, A171

\bibitem[\protect\citeauthoryear{Christie et~al.,}{Christie et~al.}{2021}]{christie_impact_2021}
Christie D.~A.,  et~al., 2021, \mn@doi [MNRAS] {10.1093/mnras/stab2027}, 506, 4500

\bibitem[\protect\citeauthoryear{Christie, Mayne, Gillard, Manners, Hébrard, Lines  \& Kohary}{Christie et~al.}{2022}]{christie_impact_2022}
Christie D.~A.,  Mayne N.~J.,  Gillard R.~M.,  Manners J.,  Hébrard E.,  Lines S.,   Kohary K.,  2022, \mn@doi [MNRAS] {10.1093/mnras/stac2763}, 517, 1407

\bibitem[\protect\citeauthoryear{Christie, Mayne, Zamyatina, Baskett, Evans-Soma, Wood  \& Kohary}{Christie et~al.}{2024}]{christie2024longitudinal}
Christie D.,  Mayne N.,  Zamyatina M.,  Baskett H.,  Evans-Soma T.,  Wood N.,   Kohary K.,  2024, \mn@doi [MNRAS] {10.1093/mnras/stae1408}, 532, 3001

\bibitem[\protect\citeauthoryear{Cloutier et~al.,}{Cloutier et~al.}{2017}]{cloutier2017characterization}
Cloutier R.,  et~al., 2017, \mn@doi [A\&A] {https://doi.org/10.1051/0004-6361/201731558}, 608, A35

\bibitem[\protect\citeauthoryear{Cooke \& Madhusudhan}{Cooke \& Madhusudhan}{2024}]{cooke2024considerations}
Cooke G.~J.,  Madhusudhan N.,  2024, \mn@doi [ApJ] {10.3847/1538-4357/ad8cda}, 977, 209

\bibitem[\protect\citeauthoryear{Drummond, Tremblin, Baraffe, Amundsen, Mayne, Venot  \& Goyal}{Drummond et~al.}{2016}]{drummond_effects_2016}
Drummond B.,  Tremblin P.,  Baraffe I.,  Amundsen D.~S.,  Mayne N.~J.,  Venot O.,   Goyal J.,  2016, \mn@doi [A\&A] {10.1051/0004-6361/201628799}, 594, A69

\bibitem[\protect\citeauthoryear{Drummond, Mayne, Baraffe, Tremblin, Manners, Amundsen, Goyal  \& Acreman}{Drummond et~al.}{2018a}]{drummond_effect_2018}
Drummond B.,  Mayne N.~J.,  Baraffe I.,  Tremblin P.,  Manners J.,  Amundsen D.~S.,  Goyal J.,   Acreman D.,  2018a, \mn@doi [A\&A] {10.1051/0004-6361/201732010}, 612, A105

\bibitem[\protect\citeauthoryear{Drummond et~al.,}{Drummond et~al.}{2018b}]{drummond_observable_2018}
Drummond B.,  et~al., 2018b, \mn@doi [ApJ] {10.3847/2041-8213/aab209}, 855, L31

\bibitem[\protect\citeauthoryear{Drummond et~al.,}{Drummond et~al.}{2020}]{drummond2020implications}
Drummond B.,  et~al., 2020, \mn@doi [A\&A] {10.1051/0004-6361/201937153}, 636, A68

\bibitem[\protect\citeauthoryear{Dymont, Yu, Ohno, Zhang, Fortney, Thorngren  \& Dickinson}{Dymont et~al.}{2022}]{dymont2022cleaning}
Dymont A.~H.,  Yu X.,  Ohno K.,  Zhang X.,  Fortney J.~J.,  Thorngren D.,   Dickinson C.,  2022, \mn@doi [ApJ] {10.3847/1538-4357/ac7f40}, 937, 90

\bibitem[\protect\citeauthoryear{Edwards}{Edwards}{1996}]{edwards1996efficient}
Edwards J.,  1996, \mn@doi [J. Atmos. Sci.] {10.1175/1520-0469(1996)053<1921:ECOIFA>2.0.CO;2}, 53, 1921

\bibitem[\protect\citeauthoryear{Edwards \& Slingo}{Edwards \& Slingo}{1996}]{edwards1996studies}
Edwards J.,  Slingo A.,  1996, \mn@doi [Q. J. R. Meteorol. Soc.] {10.1002/qj.49712253107}, 122, 689

\bibitem[\protect\citeauthoryear{Engle \& Guinan}{Engle \& Guinan}{2018}]{engle2018rotation}
Engle S.~G.,  Guinan E.~F.,  2018, \mn@doi [Research Notes of the AAS] {10.3847/2515-5172/aab1f8}, 2, 34

\bibitem[\protect\citeauthoryear{France et~al.,}{France et~al.}{2016}]{france2016muscles}
France K.,  et~al., 2016, \mn@doi [ApJ] {10.3847/0004-637X/820/2/89}, 820, 89

\bibitem[\protect\citeauthoryear{Fulton \& Petigura}{Fulton \& Petigura}{2018}]{fulton2018california}
Fulton B.~J.,  Petigura E.~A.,  2018, \mn@doi [\aj] {10.3847/1538-3881/aae828}, 156, 264

\bibitem[\protect\citeauthoryear{Guillot, Burrows, Hubbard, Lunine  \& Saumon}{Guillot et~al.}{1996}]{guillot1996giant}
Guillot T.,  Burrows A.,  Hubbard W.,  Lunine J.,   Saumon D.,  1996, \mn@doi [ApJ] {10.1086/309935}, 459, L35

\bibitem[\protect\citeauthoryear{Hindmarsh}{Hindmarsh}{1983}]{hindmarsh1983scientific}
Hindmarsh A.,  1983, Scientific Computing (IMACS Transactions on Scientific Computation vol 1) ed RS Stepleman et al

\bibitem[\protect\citeauthoryear{Holmberg \& Madhusudhan}{Holmberg \& Madhusudhan}{2024}]{holmberg2024possible}
Holmberg M.,  Madhusudhan N.,  2024, A\&A, 683, L2

\bibitem[\protect\citeauthoryear{Holton}{Holton}{1986}]{holton_dynamically_1986}
Holton J.~R.,  1986, \mn@doi [J. Geophys. Res.: Atmospheres] {10.1029/JD091iD02p02681}, 91, 2681

\bibitem[\protect\citeauthoryear{Holton \& Hakim}{Holton \& Hakim}{2013}]{holton2013introduction}
Holton J.~R.,  Hakim G.~J.,  2013, An introduction to dynamic meteorology.
Academic press

\bibitem[\protect\citeauthoryear{{Hu} \& {Damiano}}{{Hu} \& {Damiano}}{2021}]{Hu_JWST_2021}
{Hu} R.,  {Damiano} M.,  2021, {Deep Characterization of the Atmosphere of a Temperate Sub-Neptune}, JWST Proposal. Cycle 1, ID. \#2372

\bibitem[\protect\citeauthoryear{{Hu}, {Damiano}, {Scheucher}, {Kite}, {Seager}  \& {Rauer}}{{Hu} et~al.}{2021}]{Hu_solubility_nh3_2021}
{Hu} R.,  {Damiano} M.,  {Scheucher} M.,  {Kite} E.,  {Seager} S.,   {Rauer} H.,  2021, \mn@doi [\apjl] {10.3847/2041-8213/ac1f92}, \href {https://ui.adsabs.harvard.edu/abs/2021ApJ...921L...8H} {921, L8}

\bibitem[\protect\citeauthoryear{Innes \& Pierrehumbert}{Innes \& Pierrehumbert}{2022}]{innes2022atmospheric}
Innes H.,  Pierrehumbert R.~T.,  2022, \mn@doi [ApJ] {https://doi.org/10.3847/1538-4357/ac4887}, 927, 38

\bibitem[\protect\citeauthoryear{{Iro}, {B{\'e}zard}  \& {Guillot}}{{Iro} et~al.}{2005}]{Iro_2005}
{Iro} N.,  {B{\'e}zard} B.,   {Guillot} T.,  2005, \mn@doi [\aap] {10.1051/0004-6361:20048344}, \href {https://ui.adsabs.harvard.edu/abs/2005A\&A...436..719I} {436, 719}

\bibitem[\protect\citeauthoryear{Kasting, Whitmire  \& Reynolds}{Kasting et~al.}{1993}]{kasting_habitable_1993}
Kasting J.~F.,  Whitmire D.~P.,   Reynolds R.~T.,  1993, \mn@doi [Icarus] {10.1006/icar.1993.1010}, 101, 108

\bibitem[\protect\citeauthoryear{Komacek, Showman  \& Parmentier}{Komacek et~al.}{2019}]{komacek_vertical_2019}
Komacek T.~D.,  Showman A.~P.,   Parmentier V.,  2019, \mn@doi [ApJ] {10.3847/1538-4357/ab338b}, 881, 152

\bibitem[\protect\citeauthoryear{Kopparapu et~al.,}{Kopparapu et~al.}{2013}]{kopparapu_habitable_2013}
Kopparapu R.~K.,  et~al., 2013, \mn@doi [ApJ] {10.1088/0004-637X/765/2/131}, 765, 131

\bibitem[\protect\citeauthoryear{Leconte, Forget, Charnay, Wordsworth, Selsis, Millour  \& Spiga}{Leconte et~al.}{2013}]{leconte20133d}
Leconte J.,  Forget F.,  Charnay B.,  Wordsworth R.,  Selsis F.,  Millour E.,   Spiga A.,  2013, \mn@doi [A\&A] {https://doi.org/10.1051/0004-6361/201321042}, 554, A69

\bibitem[\protect\citeauthoryear{{Leconte}, {Selsis}, {Hersant}  \& {Guillot}}{{Leconte} et~al.}{2017}]{Leconte_2017}
{Leconte} J.,  {Selsis} F.,  {Hersant} F.,   {Guillot} T.,  2017, \mn@doi [\aap] {10.1051/0004-6361/201629140}, \href {https://ui.adsabs.harvard.edu/abs/2017A&A...598A..98L} {598, A98}

\bibitem[\protect\citeauthoryear{{Leconte} et~al.,}{{Leconte} et~al.}{2024}]{Leconte_2024}
{Leconte} J.,  et~al., 2024, \mn@doi [\aap] {10.1051/0004-6361/202348928}, \href {https://ui.adsabs.harvard.edu/abs/2024A&A...686A.131L} {686, A131}

\bibitem[\protect\citeauthoryear{Louden, Laughlin  \& Millholland}{Louden et~al.}{2023}]{louden2023tidal}
Louden E.~M.,  Laughlin G.~P.,   Millholland S.~C.,  2023, \mn@doi [ApJ] {10.3847/2041-8213/ad0843}, 958, L21

\bibitem[\protect\citeauthoryear{Luque \& Pallé}{Luque \& Pallé}{2022}]{luque_density_2022}
Luque R.,  Pallé E.,  2022, \mn@doi [Science] {10.1126/science.abl7164}, 377, 1211

\bibitem[\protect\citeauthoryear{Luque, Piaulet-Ghorayeb, Radica, Xue, Zhang, Bean, Samra  \& Steinrueck}{Luque et~al.}{2025}]{luque_insufficient_2025}
Luque R.,  Piaulet-Ghorayeb C.,  Radica M.,  Xue Q.,  Zhang M.,  Bean J.~L.,  Samra D.,   Steinrueck M.~E.,  2025, \mn@doi [arXiv e-prints] {10.48550/arXiv.2505.13407}

\bibitem[\protect\citeauthoryear{Madhusudhan, Nixon, Welbanks, Piette  \& Booth}{Madhusudhan et~al.}{2020}]{madhusudhan_interior_2020}
Madhusudhan N.,  Nixon M.~C.,  Welbanks L.,  Piette A. A.~A.,   Booth R.~A.,  2020, \mn@doi [ApJ] {10.3847/2041-8213/ab7229}, 891, L7

\bibitem[\protect\citeauthoryear{Madhusudhan, Sarkar, Constantinou, Holmberg, Piette  \& Moses}{Madhusudhan et~al.}{2023}]{madhusudhan_carbon-bearing_2023}
Madhusudhan N.,  Sarkar S.,  Constantinou S.,  Holmberg M.,  Piette A. A.~A.,   Moses J.~I.,  2023, \mn@doi [ApJ] {10.3847/2041-8213/acf577}, 956, L13

\bibitem[\protect\citeauthoryear{Madhusudhan, Constantinou, Holmberg, Sarkar, Piette  \& Moses}{Madhusudhan et~al.}{2025}]{madhusudhan2025new}
Madhusudhan N.,  Constantinou S.,  Holmberg M.,  Sarkar S.,  Piette A.~A.,   Moses J.~I.,  2025, \mn@doi [ApJ] {10.3847/2041-8213/adc1c8}, 983, L40

\bibitem[\protect\citeauthoryear{Marley \& Robinson}{Marley \& Robinson}{2015}]{marley_cool_2015}
Marley M.,  Robinson T.,  2015, \mn@doi [ARA\&A] {10.1146/annurev-astro-082214-122522}, 53, 279

\bibitem[\protect\citeauthoryear{Mayne et~al.,}{Mayne et~al.}{2014}]{mayne2014unified}
Mayne N.~J.,  et~al., 2014, \mn@doi [A\&A] {10.1051/0004-6361/201322174}, 561, A1

\bibitem[\protect\citeauthoryear{Mayne et~al.,}{Mayne et~al.}{2017}]{mayne2017results}
Mayne N.~J.,  et~al., 2017, \mn@doi [A\&A] {10.1051/0004-6361/201730465}, 604, A79

\bibitem[\protect\citeauthoryear{Mayne, Drummond, Debras, Jaupart, Manners, Boutle, Baraffe  \& Kohary}{Mayne et~al.}{2019}]{mayne2019limits}
Mayne N.,  Drummond B.,  Debras F.,  Jaupart E.,  Manners J.,  Boutle I.,  Baraffe I.,   Kohary K.,  2019, \mn@doi [ApJ] {10.3847/1538-4357/aaf6e9}, 871, 56

\bibitem[\protect\citeauthoryear{Misener \& Schlichting}{Misener \& Schlichting}{2021}]{misener2021cool}
Misener W.,  Schlichting H.~E.,  2021, \mn@doi [MNRAS] {10.1093/mnras/stab895}, 503, 5658

\bibitem[\protect\citeauthoryear{{Montet} et~al.,}{{Montet} et~al.}{2015}]{Montet_K2_2015}
{Montet} B.~T.,  et~al., 2015, \mn@doi [\apj] {10.1088/0004-637X/809/1/25}, \href {https://ui.adsabs.harvard.edu/abs/2015ApJ...809...25M} {809, 25}

\bibitem[\protect\citeauthoryear{{Moses} et~al.,}{{Moses} et~al.}{2011}]{Moses_2011}
{Moses} J.~I.,  et~al., 2011, \mn@doi [\apj] {10.1088/0004-637X/737/1/15}, \href {https://ui.adsabs.harvard.edu/abs/2011ApJ...737...15M} {737, 15}

\bibitem[\protect\citeauthoryear{Nixon \& Madhusudhan}{Nixon \& Madhusudhan}{2021}]{Nixon_Madhu_ocean_2021}
Nixon M.~C.,  Madhusudhan N.,  2021, \mn@doi [MNRAS] {10.1093/mnras/stab1500}, 505, 3414

\bibitem[\protect\citeauthoryear{Parmentier, Showman  \& Lian}{Parmentier et~al.}{2013}]{parmentier_3d_2013}
Parmentier V.,  Showman A.~P.,   Lian Y.,  2013, \mn@doi [A\&A] {10.1051/0004-6361/201321132}, 558, A91

\bibitem[\protect\citeauthoryear{{Piette} \& {Madhusudhan}}{{Piette} \& {Madhusudhan}}{2020}]{Piette_Madhu_2020}
{Piette} A. A.~A.,  {Madhusudhan} N.,  2020, \mn@doi [\apj] {10.3847/1538-4357/abbfb1}, \href {https://ui.adsabs.harvard.edu/abs/2020ApJ...904..154P} {904, 154}

\bibitem[\protect\citeauthoryear{Plumb \& Mahlman}{Plumb \& Mahlman}{1987}]{plumb1987zonally}
Plumb R.,  Mahlman J.,  1987, \mn@doi [J. Atmos. Sci.] {https://doi.org/10.1175/1520-0469(1987)044<0298:TZATCO>2.0.CO;2}, 44, 298

\bibitem[\protect\citeauthoryear{{Sairam} \& {Madhusudhan}}{{Sairam} \& {Madhusudhan}}{2025}]{Sairam2025arXiv250319908S}
{Sairam} L.,  {Madhusudhan} N.,  2025, \mn@doi [arXiv e-prints] {10.48550/arXiv.2503.19908}, \href {https://ui.adsabs.harvard.edu/abs/2025arXiv250319908S} {p. arXiv:2503.19908}

\bibitem[\protect\citeauthoryear{{Sarkis} et~al.,}{{Sarkis} et~al.}{2018}]{Sarkis_2018}
{Sarkis} P.,  et~al., 2018, \mn@doi [\aj] {10.3847/1538-3881/aac108}, \href {https://ui.adsabs.harvard.edu/abs/2018AJ....155..257S} {155, 257}

\bibitem[\protect\citeauthoryear{Schmidt et~al.,}{Schmidt et~al.}{2025}]{schmidt2025comprehensive}
Schmidt S.~P.,  et~al., 2025, \mn@doi [arXiv preprint arXiv:2501.18477] {https://doi.org/10.48550/arXiv.2501.18477}

\bibitem[\protect\citeauthoryear{{Seager}, {Welbanks}, {Ellerbroek}, {Bains}  \& {Petkowski}}{{Seager} et~al.}{2025}]{seager_prospects_nodate}
{Seager} S.,  {Welbanks} L.,  {Ellerbroek} L.,  {Bains} W.,   {Petkowski} J.~J.,  2025, \mn@doi [arXiv e-prints] {10.48550/arXiv.2504.12946}, \href {https://ui.adsabs.harvard.edu/abs/2025arXiv250412946S} {p. arXiv:2504.12946}

\bibitem[\protect\citeauthoryear{Shorttle, Jordan, Nicholls, Lichtenberg  \& Bower}{Shorttle et~al.}{2024}]{shorttle_distinguishing_2024}
Shorttle O.,  Jordan S.,  Nicholls H.,  Lichtenberg T.,   Bower D.~J.,  2024, \mn@doi [ApJ] {10.3847/2041-8213/ad206e}, 962, L8

\bibitem[\protect\citeauthoryear{Showman \& Polvani}{Showman \& Polvani}{2011}]{showman_equatorial_2011}
Showman A.~P.,  Polvani L.~M.,  2011, \mn@doi [ApJ] {10.1088/0004-637X/738/1/71}, 738, 71

\bibitem[\protect\citeauthoryear{{Showman}, {Wordsworth}, {Merlis}  \& {Kaspi}}{{Showman} et~al.}{2013}]{Showman_circulation_2013}
{Showman} A.~P.,  {Wordsworth} R.~D.,  {Merlis} T.~M.,   {Kaspi} Y.,  2013, in {Mackwell} S.~J.,  {Simon-Miller} A.~A.,  {Harder} J.~W.,   {Bullock} M.~A.,  eds, , Comparative Climatology of Terrestrial Planets.
pp 277--327, \mn@doi{10.2458/azu_uapress_9780816530595-ch012}

\bibitem[\protect\citeauthoryear{Showman, Tan  \& Parmentier}{Showman et~al.}{2020}]{showman_atmospheric_2020}
Showman A.~P.,  Tan X.,   Parmentier V.,  2020, \mn@doi [Space Sci. Rev] {10.1007/s11214-020-00758-8}, 216, 139

\bibitem[\protect\citeauthoryear{{Taylor}}{{Taylor}}{2025}]{taylor_are_2025}
{Taylor} J.,  2025, \mn@doi [arXiv e-prints] {10.48550/arXiv.2504.15916}, \href {https://ui.adsabs.harvard.edu/abs/2025arXiv250415916T} {p. arXiv:2504.15916}

\bibitem[\protect\citeauthoryear{Tremblin, Amundsen, Mourier, Baraffe, Chabrier, Drummond, Homeier  \& Venot}{Tremblin et~al.}{2015}]{tremblin_fingering_2015}
Tremblin P.,  Amundsen D.~S.,  Mourier P.,  Baraffe I.,  Chabrier G.,  Drummond B.,  Homeier D.,   Venot O.,  2015, \mn@doi [ApJ] {10.1088/2041-8205/804/1/L17}, 804, L17

\bibitem[\protect\citeauthoryear{Tremblin, Amundsen, Chabrier, Baraffe, Drummond, Hinkley, Mourier  \& Venot}{Tremblin et~al.}{2016}]{tremblin2016cloudless}
Tremblin P.,  Amundsen D.~S.,  Chabrier G.,  Baraffe I.,  Drummond B.,  Hinkley S.,  Mourier P.,   Venot O.,  2016, \mn@doi [\apjl] {10.3847/2041-8205/817/2/L19}, 817, L19

\bibitem[\protect\citeauthoryear{Tsai, Innes, Lichtenberg, Taylor, Malik, Chubb  \& Pierrehumbert}{Tsai et~al.}{2021}]{tsai_inferring_2021}
Tsai S.-M.,  Innes H.,  Lichtenberg T.,  Taylor J.,  Malik M.,  Chubb K.,   Pierrehumbert R.,  2021, \mn@doi [ApJ] {10.3847/2041-8213/ac399a}, 922, L27

\bibitem[\protect\citeauthoryear{Venot, Hébrard, Agúndez, Dobrijevic, Selsis, Hersant, Iro  \& Bounaceur}{Venot et~al.}{2012}]{venot_chemical_2012}
Venot O.,  Hébrard E.,  Agúndez M.,  Dobrijevic M.,  Selsis F.,  Hersant F.,  Iro N.,   Bounaceur R.,  2012, \mn@doi [A\&A] {10.1051/0004-6361/201219310}, 546, A43

\bibitem[\protect\citeauthoryear{Venot, Bounaceur, Dobrijevic, H{\'e}brard, Cavali{\'e}, Tremblin, Drummond  \& Charnay}{Venot et~al.}{2019}]{venot2019reduced}
Venot O.,  Bounaceur R.,  Dobrijevic M.,  H{\'e}brard E.,  Cavali{\'e} T.,  Tremblin P.,  Drummond B.,   Charnay B.,  2019, \mn@doi [A\&A] {10.1051/0004-6361/201834861}, 624, A58

\bibitem[\protect\citeauthoryear{Visscher}{Visscher}{2012}]{visscher_chemical_2012}
Visscher C.,  2012, \mn@doi [ApJ] {10.1088/0004-637X/757/1/5}, 757, 5

\bibitem[\protect\citeauthoryear{{Visscher} \& {Moses}}{{Visscher} \& {Moses}}{2011}]{Visscher_2011}
{Visscher} C.,  {Moses} J.~I.,  2011, \mn@doi [\apj] {10.1088/0004-637X/738/1/72}, \href {https://ui.adsabs.harvard.edu/abs/2011ApJ...738...72V} {738, 72}

\bibitem[\protect\citeauthoryear{Wang \& Wordsworth}{Wang \& Wordsworth}{2020}]{wang2020extremely}
Wang H.,  Wordsworth R.,  2020, \mn@doi [ApJ] {10.3847/1538-4357/ab6dcc}, 891, 7

\bibitem[\protect\citeauthoryear{{Welbanks} et~al.,}{{Welbanks} et~al.}{2025}]{welbanks_challenges_2025}
{Welbanks} L.,  et~al., 2025, \mn@doi [arXiv e-prints] {10.48550/arXiv.2504.21788}, \href {https://ui.adsabs.harvard.edu/abs/2025arXiv250421788W} {p. arXiv:2504.21788}

\bibitem[\protect\citeauthoryear{Wogan, Batalha, Zahnle, Krissansen-Totton, Tsai  \& Hu}{Wogan et~al.}{2024}]{wogan2024jwst}
Wogan N.~F.,  Batalha N.~E.,  Zahnle K.~J.,  Krissansen-Totton J.,  Tsai S.-M.,   Hu R.,  2024, \mn@doi [ApJ] {10.3847/2041-8213/ad2616}, 963, L7

\bibitem[\protect\citeauthoryear{Wood et~al.,}{Wood et~al.}{2014}]{wood2014inherently}
Wood N.,  et~al., 2014, \mn@doi [Q. J. R. Meteorol. Soc.] {10.1002/qj.2235}, 140, 1505

\bibitem[\protect\citeauthoryear{Yu et~al.,}{Yu et~al.}{2021a}]{yu2021haze}
Yu X.,  et~al., 2021a, \mn@doi [Nature Astron.] {10.1038/s41550-021-01375-3}, 5, 822

\bibitem[\protect\citeauthoryear{Yu, Moses, Fortney  \& Zhang}{Yu et~al.}{2021b}]{yu__how_2021}
Yu X.,  Moses J.~I.,  Fortney J.~J.,   Zhang X.,  2021b, \mn@doi [ApJ] {10.3847/1538-4357/abfdc7}, 914, 38

\bibitem[\protect\citeauthoryear{Yurchenko, Mellor, Freedman  \& Tennyson}{Yurchenko et~al.}{2020}]{yurchenko2020exomol}
Yurchenko S.,  Mellor T.~M.,  Freedman R.~S.,   Tennyson J.,  2020, \mn@doi [MNRAS] {10.1093/mnras/staa1874}, 496, 5282

\bibitem[\protect\citeauthoryear{Yurchenko, Owens, Kefala  \& Tennyson}{Yurchenko et~al.}{2024}]{yurchenko_exomol_2024}
Yurchenko S.~N.,  Owens A.,  Kefala K.,   Tennyson J.,  2024, \mn@doi [MNRAS] {10.1093/mnras/stae148}, 528, 3719

\bibitem[\protect\citeauthoryear{Zahnle \& Marley}{Zahnle \& Marley}{2014}]{zahnle_methane_2014}
Zahnle K.~J.,  Marley M.~S.,  2014, \mn@doi [ApJ] {10.1088/0004-637X/797/1/41}, 797, 41

\bibitem[\protect\citeauthoryear{Zamyatina et~al.,}{Zamyatina et~al.}{2022}]{zamyatina_observability_2022}
Zamyatina M.,  et~al., 2022, \mn@doi [MNRAS] {10.1093/mnras/stac3432}, 519, 3129

\bibitem[\protect\citeauthoryear{Zamyatina et~al.,}{Zamyatina et~al.}{2024}]{zamyatina_quenching-driven_2024}
Zamyatina M.,  et~al., 2024, \mn@doi [MNRAS] {10.1093/mnras/stae600}, 529, 1776

\bibitem[\protect\citeauthoryear{{Zeng} et~al.,}{{Zeng} et~al.}{2019}]{Zeng_planet_distribution_2019}
{Zeng} L.,  et~al., 2019, \mn@doi [PNAS] {10.1073/pnas.1812905116}, \href {https://ui.adsabs.harvard.edu/abs/2019PNAS..116.9723Z} {116, 9723}

\bibitem[\protect\citeauthoryear{Zhang \& Showman}{Zhang \& Showman}{2018a}]{zhang_global-mean_2018}
Zhang X.,  Showman A.~P.,  2018a, \mn@doi [ApJ] {10.3847/1538-4357/aada85}, 866, 1

\bibitem[\protect\citeauthoryear{{Zhang} \& {Showman}}{{Zhang} \& {Showman}}{2018b}]{Zhang_2018_II}
{Zhang} X.,  {Showman} A.~P.,  2018b, \mn@doi [\apj] {10.3847/1538-4357/aada7c}, \href {https://ui.adsabs.harvard.edu/abs/2018ApJ...866....2Z} {866, 2}

\makeatother
\end{thebibliography}



\newpage
\appendix

\section{Methods to obtain stability and conservation for UM}\label{stability}

\begin{figure*}
\centering
\includegraphics[width=1.8\columnwidth]{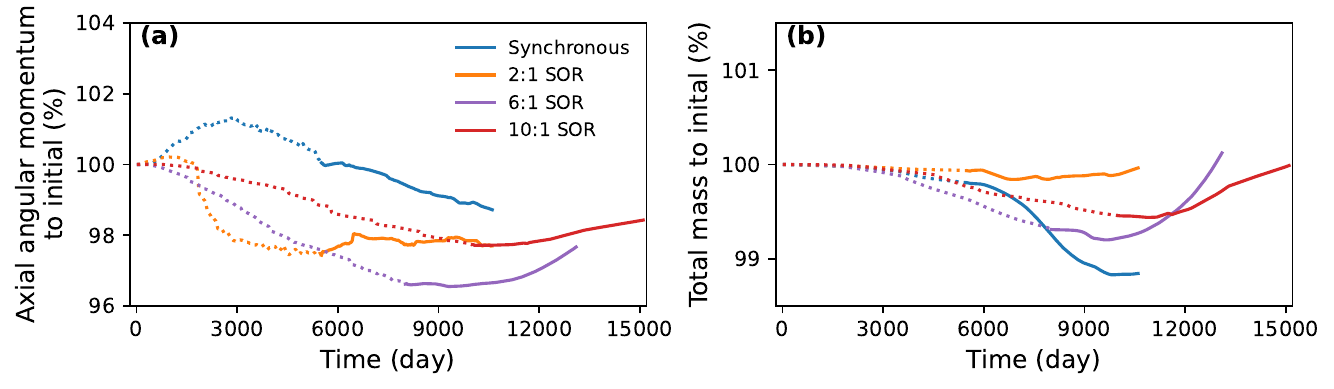}
\caption{The evolution of axial angular momentum (a) and total mass (b) with time in the simulations. Dotted lines indicate the fixed abundance runs, while the solid lines indicate the kinetics runs. Different coloured lines are from different simulations.}
\label{fig:conservation}
\end{figure*}

The model employs three methods to obtain stability: longitudinal filtering, vertical damping at the lower and upper boundaries, and bottom drag. We employ a first-order longitudinal filtering of the horizontal wind $\mathbf{u_h}$:
\begin{equation}
    (\frac{\partial \mathbf{u_h}}{\partial t}) = K_{\mathrm{eff}}\nabla^2_\lambda \mathbf{u_h},
\end{equation}
where $\lambda$ is the longitude. The effective longitudinal diffusion coefficient is
\begin{equation}
    K_{\mathrm{eff}} \sim K \frac{r^2 \cos^2 \phi (\Delta \lambda)^2}{\Delta t_{\mathrm{dyn}}^2},
\end{equation}
where $\phi$ and $r$ are the latitude and radius, and $\Delta t_{\mathrm{dyn}}$ is the dynamical timescale. The filtering constant $K$ is given by
\begin{equation}
    K = \frac{1}{4} \left( 1 - e^{-\frac{1}{t_K}} \right).
\end{equation}
In this study, we set $t_K=4$, which is the largest value that ensures the stability of the model. 

Damping is implicitly applied to the vertical velocity in the uppermost and lowest 6 layers (corresponding to pressure levels above 10 Pa and pressure levels below $\sim$140~bar) in the vertical direction to limit the vertical wind speed and prevent the loss of atmospheric mass from the bottom. The damping is parameterized as
\begin{equation}
    w^{t+\Delta t} = w^t - R_w\Delta t w^{t+\Delta t},
\end{equation}
where $w^t$ and $w^{t+\Delta t}$ are the vertical velocities at the current and next timestep and $\Delta t$ is the length of the timestep. The damping coefficient $R_w$ is given by 
\begin{equation}
R_w = 
\left\{
\begin{array}{ll}
0.1 \sin^2 \left[ \frac{\pi}{2} 
\left( \frac{\eta - 0.9}{0.1} \right) \right], & \eta \geq 0.9 \\[8pt]
0, & 0.1 < \eta < 0.9 \\[8pt]
4 \sin^2 \left[ \frac{\pi}{2} 
\left( \frac{0.1 - \eta}{0.1} \right) \right], & \eta \leq 0.1
\end{array}
\right.
\end{equation}
where $\eta=z/z_\mathrm{top}$ is the non-dimensional height. Damping at the upper boundary is a common setting in previous simulations of hydrogen-rich planets using UM \citep[e.g.,][]{mayne2014unified, mayne2017results, christie2024longitudinal}. In contrast, bottom damping is less commonly used but is necessary in our simulations due to the presence of a convective zone at the bottom boundary layer (discussed in section \ref{subsec:thermal}). This bottom convective zone increases the model's susceptibility to instability and leads to significant atmospheric mass loss ($\sim$10$\%$) in long simulations ($>$ 2000 days) if the bottom damping is not applied. 

A Rayleigh drag is employed when the pressure is larger than 140~bar to limit the horizontal wind speed in the bottom boundary:
\begin{equation}
    \mathbf{F_u} = -\frac{\mathbf{u}}{\tau_\mathrm{fric}},
\end{equation}
where $\tau_\mathrm{fric}$ is the friction timescale. The friction timescale is parameterized as 
\begin{equation}
    \frac{1}{\tau_\mathrm{fric}} = \frac{1}{\tau_\mathrm{fric,f}}\mathrm{max}\{{0,\frac{\sigma-0.7}{0.3}}\},
\end{equation}
where $\tau_\mathrm{fric,f} = $1 day, and $\sigma=\frac{p}{p_\mathrm{bot}}$ is the non-dimensional pressure.

The above methods can ensure the model runs stably for at least ten thousand days, with the angular momentum loss within 3 percent and the total mass loss within 1.5 percent (Fig.~\ref{fig:conservation}).

\section{Model convergence}\label{convergence}

\begin{figure*}
\centering
\includegraphics[width=1.8\columnwidth]{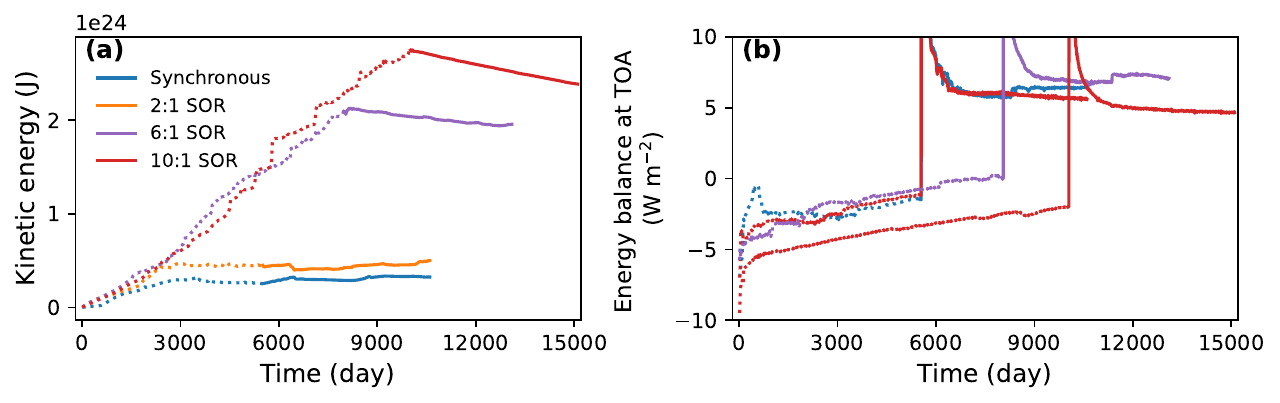}
\caption{The evolution of kinetic energy (a) and the net radiation at the TOA (b) with time in the simulations. Dotted lines indicate the fixed abundance runs, while the solid lines indicate the kinetics runs. Different coloured lines are from different simulations.}
\label{fig:convergence}
\end{figure*}

\begin{figure*}
\centering
\includegraphics[width=2\columnwidth]{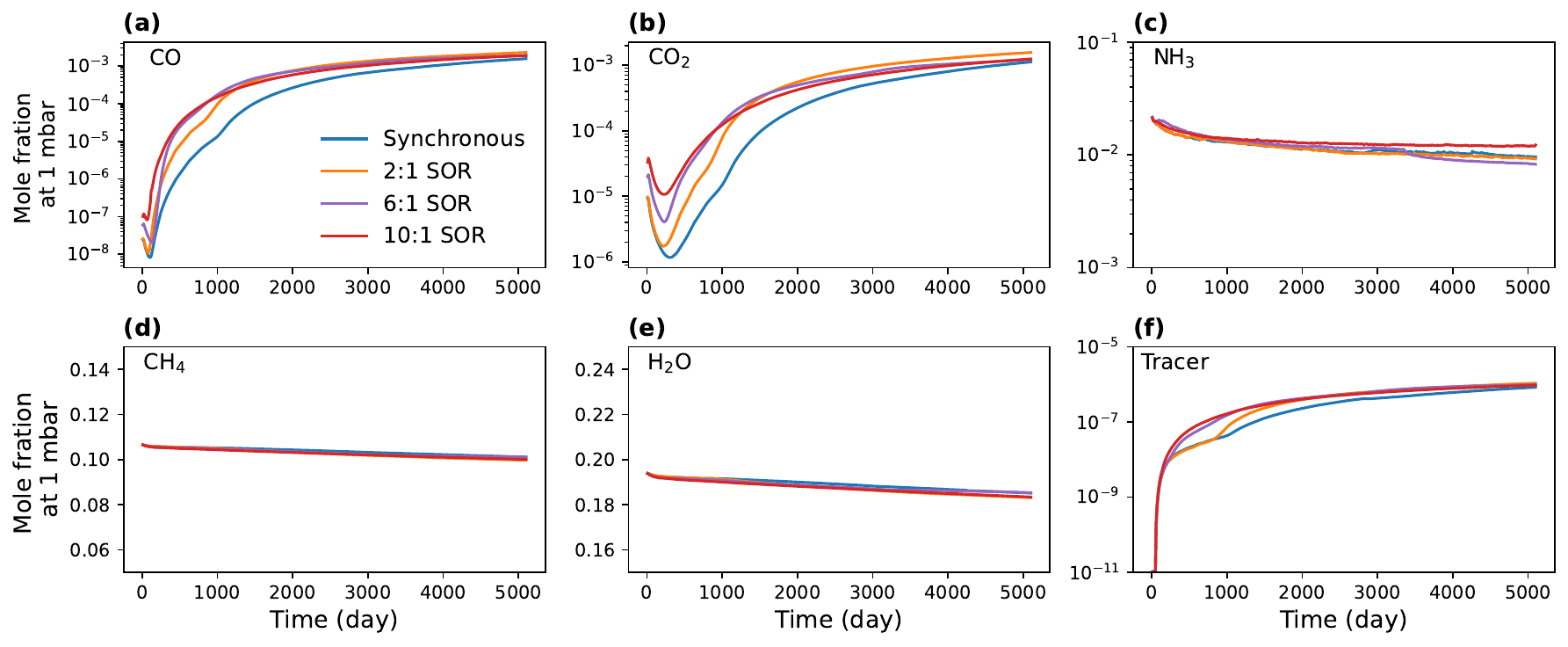}
\caption{Evolution of the global mean mole fraction of chemical species ((a)--(e)) and the passive tracer mass mixing ratio (f) at 0.001~bar through time in the kinetics runs. Different coloured lines are from different simulations.}
\label{fig:evolution}
\end{figure*}
To check the convergence of the model, we analyse the evolution of kinetic energy (Fig.~\ref{fig:convergence}(a)). In the fixed abundance runs, simulations assuming synchronous and 2:1 SOR converge after $\sim$3000 days, and the kinetic energy remains steady after chemical kinetics is coupled. The kinetic energy of simulations of 6:1 SOR and 10:1 SOR still evolves after running for 8000 and 10,000 days, but then recedes after chemical kinetics is coupled. These results imply that simulations with a faster rotation period require a longer time to converge. 

The energy budget at the top of the atmosphere (TOA) is presented in Fig.~\ref{fig:convergence}(b). By the end of the fixed abundance runs, the energy imbalance at TOA is within 3 W\,m$^{-2}$. At the end of the kinetics runs, a net imbalance of 6 W\,m$^{-2}$ remains. However, this imbalance is relatively small, as it accounts for less than 2\% of the incoming stellar radiation and is comparable to the internal heat source added at the bottom, 3.7 W\,m$^{-2}$.

The kinetics runs have been run for 5100 days, and this duration is long enough for the abundances of chemical species and passive tracer at the photosphere to reach quasi-steady states, as shown in Fig.~\ref{fig:evolution}. 

\section{1D simulations using ATMO}\label{1D}

\begin{figure*}
\centering
\includegraphics[width=1.8\columnwidth]{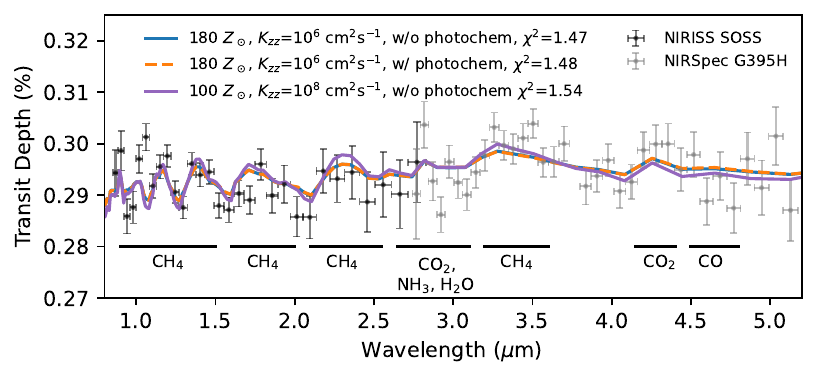}
\caption{Synthetic transmission spectra of K2-18b compared to JWST NIRISS and NIRSpec data in \citet[][fig. 3]{madhusudhan_carbon-bearing_2023}. Different lines show the synthetic spectra calculated from different 1D chemical kinetics simulations. Except for the two best-agreement simulations (blue and orange), the transmission spectrum calculated from 100 $Z_\odot$ and 10$^6$ cm$^2$\,s$^{-1}$ $K_{zz}$ simulation (purple) is also provided for comparison. The dominant molecules that contribute to the features in different spectral regions are labelled.}
\label{STS_1D}
\end{figure*}

\begin{figure*}
\centering
\includegraphics[width=1.9\columnwidth]{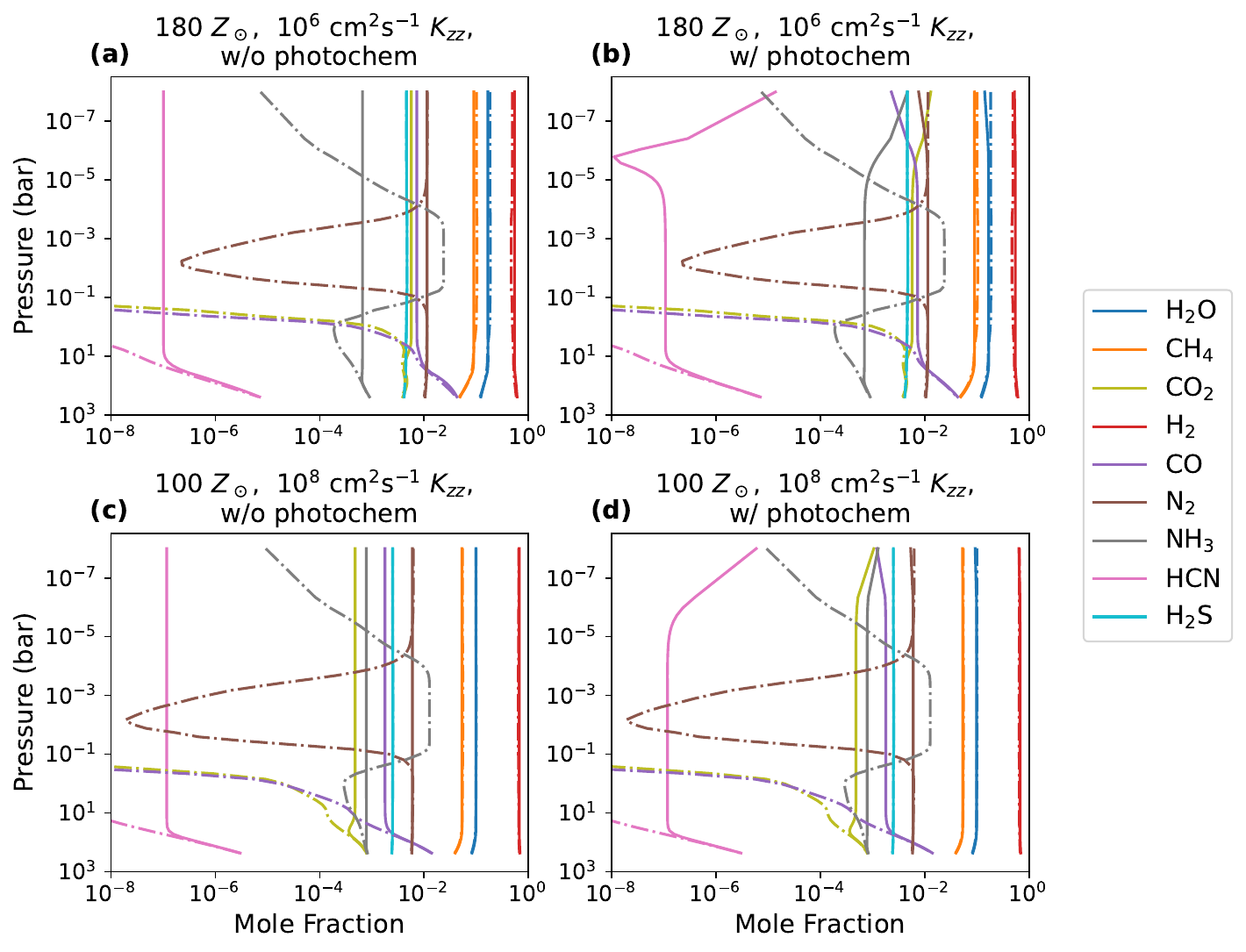}
\caption{Atmospheric composition profiles computed with \texttt{ATMO}. Except for the two best-agreement simulations with 180 $Z_\odot$ and 10$^6$ cm$^2$\,s$^{-1}$ $K_{zz}$ ((a) and (b)), the atmospheric composition profiles from 100 $Z_\odot$ and 10$^8$ cm$^2$\,s$^{-1}$ $K_{zz}$ simulations ((c) and (d)) are also provided for comparison. The dashed lines are results from chemical equilibrium calculations.}
\label{ATMO_mole}
\end{figure*}

\begin{table}
	\centering
	\caption{The $\chi^2$ value each has 64 degrees of freedom. w/ stands for `with,' and w/o stands for `without'.}
	\label{ATMO_sum}
	\begin{tabular}{lccr} 
		\hline
		Metallicity ($\times$$Z_\odot$) & $K_{zz}$ (cm$^2$\,s$^{-1}$) & Photochemistry & $\chi^2$\\
		\hline
		180 & 10$^6$ & w/o & 1.47 \\
            180 & 10$^8$ & w/o & 1.49 \\
            180 & 10$^6$ & w/ & 1.48 \\
            180 & 10$^8$ & w/ & 1.49 \\
            100 & 10$^6$ & w/o & 1.55 \\
            100 & 10$^8$ & w/o & 1.54 \\
            100 & 10$^6$ & w/ & 1.55 \\
            100 & 10$^8$ & w/ & 1.53 \\
		\hline
	\end{tabular}
\end{table}

To generate an initial temperature profile and constrain the atmospheric composition for the 3D simulations, we first conduct a series of 1D simulations using the radiative-convective equilibrium model \texttt{ATMO} \citep{tremblin_fingering_2015,tremblin2016cloudless,drummond_effects_2016}. \texttt{ATMO} uses the original chemical network developed by \cite{venot_chemical_2012}.

We begin by generating temperature profiles with opacities for 100 and 180 times solar metallicity ($Z_\odot$) under chemical equilibrium conditions, assuming a solar C/O ratio (0.55), a geometry factor of 0.25, and an internal temperature of 60~K. We include the opacity of CH$_4$, H$_2$O, CO, CO$_2$, NH$_3$, Na, K, Li, and Rb, and the collision-induced absorption of H$_2$-H$_2$ and H$_2$-He in the simulations. The temperature profiles are then used to perform chemical kinetics and photochemical kinetics simulations. Since the spectrum of K2-18 has not been measured, we instead use the UV spectrum of GJ 176 for the photochemical calculations as discussed in section \ref{subsec:K2-18b set up}. 

In these 1D simulations, we explore two values for the eddy-diffusion coefficient ($K_{zz}$): 10$^6$ cm$^2$\,s$^{-1}$ and 10$^8$ cm$^2$\,s$^{-1}$. The choice of metallicity and $K_{zz}$ values is based on prior 1D studies of K2-18b \citep[e.g.,][]{blain_1d_2021, bezard_methane_2022, wogan2024jwst}. Clouds and hazes are excluded from these simulations.

Next, we compute synthetic transmission spectra from the simulation results and compare them to JWST observations from \citet{madhusudhan_carbon-bearing_2023} to identify the optimal parameters for 3D simulations. The $\chi^2$ values from the 1D simulations are summarized in Table \ref{ATMO_sum}. The results show that simulations with 180 $Z_\odot$ and $K_{zz}$ = 10$^6$ cm$^2$\,s$^{-1}$, excluding photochemistry, provide the best agreement with the observational data ($\chi^2$ = 1.47). Therefore, we use the molecular abundances from this simulation for the 3D fixed abundance runs and 180 $Z_\odot$ for the 3D chemical kinetics runs. 

Additionally, we perform 1D simulations including photochemistry using 100 $Z_\odot$ and 180 $Z_\odot$, finding that the results are consistent with those from the photochemistry-exclusive simulations. Among these, the 10$^6$ cm$^2$\,s$^{-1}$ $K_{zz}$ case still provides a good fit ($\chi^2$ = 1.48). The synthetic transmission spectra for the two best-agreement models (180 $Z_\odot$ and 10$^6$ cm$^2$\,s$^{-1}$ $K_{zz}$ with or without photochemistry) are shown in Fig.~\ref{STS_1D}, and the atmospheric composition profiles from both models are displayed in Fig.~\ref{ATMO_mole}.

Atmospheric compositions under chemical equilibrium are also shown as dashed lines in Fig.~\ref{ATMO_mole}. 

\begin{figure*}
\centering
\includegraphics[width=2\columnwidth]{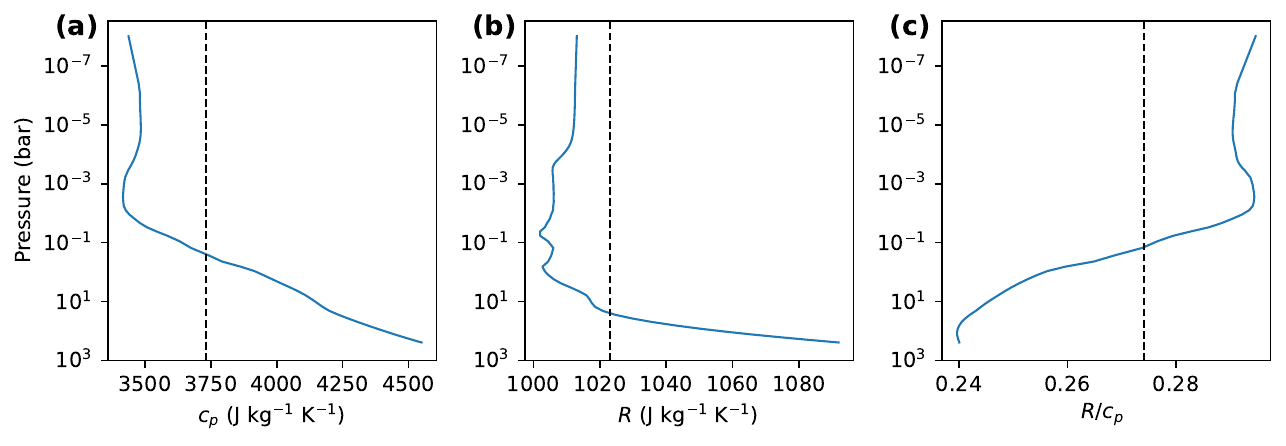}
\caption{Vertical profiles of $c_p$, $R$, and $R/c_p$ in the 180 $Z_\odot$ and 10$^6$ cm$^2$\,s$^{-1}$ $K_{zz}$ \texttt{ATMO} chemical kinetics simulation. The dashed black lines are the values used in the 3D simulations, 3733 J~kg$^{-1}$~K$^{-1}$, 1023 J~kg$^{-1}$~K$^{-1}$, and 0.274, for $c_p$, $R$, and $R/c_p$,  respectively.}
\label{Cp_R}
\end{figure*}

As mentioned in section~\ref{subsec:K2-18b set up}, specific gas constant $R$ and specific heat capacity $c_p$ used in the 3D simulations are derived from the best-agreement 1D \texttt{ATMO} simulation. In Fig. \ref{Cp_R}, we show how these two values and the dry adiabatic lapse rate ($R/c_p$) change with pressure.

\section{Derivation of the zonal- and temporal-mean zonal momentum equation}\label{derivation}

We derive the zonal- and temporal-mean zonal momentum equation from \citet[][equation (C.17)]{mayne2017results}:
\begin{align}\label{orginal momentum equation}
\frac{\partial (\rho u)}{\partial t} = 
& - \frac{1}{r \cos \phi} \frac{\partial (\rho u^2)}{\partial \lambda}
 - \frac{1}{r \cos^2 \phi} \frac{\partial (\rho uv \cos^2 \phi)}{\partial \phi} - \frac{1}{r^3} \frac{\partial (\rho u w r^3)}{\partial r} \nonumber\\
& - \frac{1}{r \cos \phi} \frac{\partial p}{\partial \lambda}
+ 2 \Omega \rho v \sin \phi - 2 \Omega \rho w \cos \phi + \rho G_\lambda.    
\end{align}

The equation above is derived by combining the zonal momentum equation in spherical coordinates for a non-hydrostatic deep atmosphere \citep{mayne2014unified}:
\begin{align} 
\frac{\mathrm{D} u}{\mathrm{D}t} =& \frac{\partial u}{\partial t}
+ \frac{u}{r \cos \phi} \frac{\partial u}{\partial \lambda}
+ \frac{v}{r} \frac{\partial u}{\partial \phi}
+ w \frac{\partial u}{\partial r} \nonumber\\
=&-\frac{uw}{r} + \frac{uv}{r} \tan \phi + 2\Omega v \sin \phi - 2\Omega w \cos \phi \nonumber\\
&- \frac{1}{r \cos \phi} \frac{1}{\rho} \frac{\partial p}{\partial \lambda} + G_{\lambda}, \end{align}
together with the mass continuity equation:
\begin{equation} 
\frac{\partial \rho}{\partial t} + \nabla \cdot (\rho \mathbf{u}) = 0. 
\end{equation}
A detailed deviation can be found in \citet[][Appendix C]{mayne2017results}.

We apply a zonal and temporal average to equation (\ref{orginal momentum equation}). By doing this, $\frac{\partial}{\partial \lambda}$ terms become zero and the wind terms can be decomposed into a zonal- and temporal-mean and derivations: $[\overline{\rho u}]=[\overline{\rho}][\overline{u}]+[\overline{\rho}^*\overline{u}^*]+[\overline{\rho'u'}]$, $[\overline{\rho uv}]=[\overline{\rho v}][\overline{u}]+[\overline{\rho v}^*\overline{u}^*]+[\overline{(\rho v)'u'}]$, and $[\overline{\rho wu}]=[\overline{\rho w}][\overline{u}]+[\overline{\rho w}^*\overline{u}^*]+[\overline{(\rho w)'u'}$)], where overbars (primes) and brackets (asterisks) denote temporal and zonal averages (deviations), respectively. This yields:
\begin{align}
\frac{\partial([\overline{\rho}][\overline{u}])}{\partial t} = 
& -\frac{1}{r\cos^2\phi}\{\frac{\partial([\overline{\rho v}][\overline{u}]\cos^2\phi)}{\partial \phi}+\frac{\partial([\overline{\rho v}^*\overline{u}^*]\cos^2\phi)}{\partial \phi} \nonumber \\
& + \frac{\partial([\overline{(\rho v)'u'}]\cos^2\phi)}{\partial \phi}\} -\frac{1}{r^3}\{\frac{\partial([\overline{\rho w}][\overline{u}]r^3)}{\partial r} \nonumber\\
&+\frac{\partial([\overline{\rho w}^*\overline{u}^*]r^3)}{\partial r} + \frac{\partial([\overline{(\rho w)'u'}]r^3)}{\partial r}\} \nonumber\\
& +2\Omega[\overline{\rho v}]\sin\phi - 2\Omega[\overline{\rho w}]\cos \phi + [\overline{\rho G_\lambda}] \nonumber\\
&-\frac{\partial([\overline{\rho'u'}])}{\partial t}
-\frac{\partial([\overline{\rho}^*\overline{u}^*])}{\partial t} .
\end{align}
Rearranging the above equation yields equation (\ref{momentum equation}) in section \ref{subsec:circulation}.

\section{Derivation of the eddy mean interaction equation for passive tracer mass mixing ratio}\label{derivation_tracer}

The tracer mass mixing ratio equation without source/sink term in the spherical coordinate is:
\begin{equation}\label{tracer_equation}
\frac{\partial q}{\partial t} = -\frac{u}{r\mathrm{cos}\phi}\frac{\partial q}{\partial \lambda}-\frac{v}{r}\frac{\partial q}{\partial \phi} -w\frac{\partial q}{\partial r} .
\end{equation}
The mass continuity equation in the spherical coordinate is:
\begin{equation}\label{mass_continuity}
\frac{\partial \rho}{\partial t} = -\frac{1}{r\mathrm{cos}\phi}\frac{\partial (\rho u)}{\partial \lambda}-\frac{1}{r\cos \phi}\frac{\partial (\rho v \cos \phi)}{\partial \phi} -\frac{1}{r^2}\frac{\partial (\rho w r^2)}{\partial r} .
\end{equation}
Multiplying equation (\ref{tracer_equation}) with $\rho$ and adding to equation (\ref{mass_continuity}) multiplying by $q$, yields:
\begin{align}\label{pq}
\rho\frac{\partial q}{\partial t}+q\frac{\partial\rho}{\partial t} =&
-\frac{1}{r\cos \phi}\{\rho u\frac{\partial q}{\partial \lambda}+q\frac{\partial (\rho u)}{\partial \lambda}\} \nonumber\\
&-\frac{1}{r\cos\phi}\{\rho v \cos\phi\frac{\partial q}{\partial \phi}+q \frac{\partial(\rho v \cos \phi)}{\partial \phi}\} \nonumber\\
&-\frac{1}{r^2}\{wr^2\frac{\partial q}{\partial r}+q\frac{\partial(\rho wr^2)}{\partial r}\}.
\end{align}
Using the product rule, equation (\ref{pq}) can be written as:
\begin{equation}\label{pq_t}
\frac{\partial (\rho q)}{\partial t} = -\frac{1}{r\cos \phi}\frac{\partial (\rho uq)}{\partial \lambda}-\frac{1}{r\cos\phi}\frac{\partial(\rho vq \cos \phi)}{\partial \phi}-\frac{1}{r^2}\frac{\partial(\rho wqr^2)}{\partial r}.    
\end{equation}
Then we apply a zonal and temporal average to equation (\ref{pq_t}). By doing this, $\frac{\partial}{\partial \lambda}$ becomes zero, and the tracer terms can be decomposed into a zonal- and temporal-mean and derivations: $[\overline{\rho q}]=[\overline{\rho}][\overline{q}]+[\overline{\rho}^*\overline{q}^*]+[\overline{\rho'q'}]$, $[\overline{\rho vq}]=[\overline{\rho v}][\overline{q}]+[\overline{\rho v}^*\overline{q}^*]+[\overline{(\rho v)'q'}$], and $[\overline{\rho wq}]=[\overline{\rho w}][\overline{q}]+[\overline{\rho w}^*\overline{q}^*]+[\overline{(\rho w)'q'}]$. This yields the mean and eddy interaction equation for passive tracer mass mixing ratio:
\begin{align}
\frac{\partial{([\overline{\rho}]}[\overline{q}])}{\partial{t}} = 
&-\frac{1}{r\cos\phi}\{\frac{\partial([\overline{\rho v}][\overline{q}]\cos\phi)}{\partial \phi}+\frac{\partial({[\overline{(\rho v)'q'}]\cos\phi)}}{\partial \phi} \nonumber\\
&+\frac{\partial([\overline{\rho v}^*\overline{q}^*]\cos\phi)}{\partial \phi}\} -\frac{1}{r^2}\{\frac{\partial([\overline{\rho w}][\overline{q}]r^2)}{\partial t} \nonumber\\
&+\frac{\partial([\overline{(\rho w)'q'}]r^2)}{\partial r}+\frac{\partial([\overline{\rho w}^*\overline{q}^*]r^2)}{\partial r}\}   \nonumber\\
& - \frac{\partial([\overline{\rho'q'}])}{\partial t} -\frac{\partial([\overline{\rho}^*\overline{q}^*])}{\partial t}. 
\end{align}


\bsp	
\label{lastpage}
\end{document}